\documentclass[acmsmall]{acmart}

\PassOptionsToPackage{prologue,table}{xcolor}
\PassOptionsToPackage{hyphens}{url}

\usepackage{hyperref}
\usepackage{breakcites}
\usepackage{booktabs}
\usepackage{multirow}
\usepackage{xurl}
\usepackage{todonotes}
\usepackage{comment}
\usepackage{rotating}
\usepackage{listings, xcolor}
\usepackage{orcidlink}
\usepackage{pifont}
\usepackage{caption}
\usepackage{subcaption}
\usepackage{array}
\usepackage{colortbl}
\usepackage{hhline}
\usepackage[ruled,linesnumbered,noend]{algorithm2e}
\usepackage{enumitem}
\usepackage{tabularx}
\usepackage{hyphenat}
\usepackage{makecell}
\usepackage{wrapfig}

\long\def\longcaption#1#2{\centering\begin{minipage}{#1}\vspace{-0.8\baselineskip}\scriptsize\noindent#2\end{minipage}}
\long\def\longcaptionfig#1#2{\centering\begin{minipage}{#1}\vspace{0.3\baselineskip}\scriptsize\noindent#2\end{minipage}}

\newcommand{\highlight}[2]{%
    \vspace{0.2\baselineskip}%
    \colorlet{currentcolor}{.}%
    {\color{#1}%biolinum
    \noindent\fbox{\parbox{0.97\linewidth}{\color{currentcolor}#2}}}%
    \vspace{0.3\baselineskip}
}

\newcommand*{\belowrulesepcolor}[1]{% 
  \noalign{% 
    \kern-\belowrulesep 
    \begingroup 
      \color{#1}% 
      \hrule height\belowrulesep 
    \endgroup 
  }%
} 
\newcommand*{\aboverulesepcolor}[1]{% 
  \noalign{% 
    \begingroup 
      \color{#1}% 
      \hrule height\aboverulesep 
    \endgroup 
    \kern-\aboverulesep 
  }%
} 
\newcommand{\Ideal}{\emph{Retrospective Training - Retrospective Testing (R-R)}\xspace}

\newcommand{\Real}{\emph{Retrospective Training - Perspective Testing (R-P)}\xspace}

\newcommand{\RealDataset}{Perspective test set\xspace}
\newcommand{\IdealTestDataset}{retrospective}

\newcommand{\RealTestDataset}{Perspective}
\newcommand{\IdealShort}{\emph{R-R}\xspace}

\newcommand{\RealShort}{\emph{R-P}\xspace}

\setcopyright{cc}
\setcctype{by}
\acmDOI{10.1145/3715731}
\acmYear{2025}
\acmJournal{PACMSE}
\acmVolume{2}
\acmNumber{FSE}
\acmArticle{FSE016}
\acmMonth{7}
\received{2024-09-11}
\received[accepted]{2025-01-14}

\author{Ranindya Paramitha}
\orcid{0000-0002-6682-4243}
\affiliation{%
  \institution{University of Trento}
  \city{Trento}
  \country{Italy}
}
\email{ranindya.paramitha@unitn.it}

\author{Yuan Feng}
\orcid{0009-0007-0641-7260}
\affiliation{%
  \institution{University of Trento}
  \city{Trento}
  \country{Italy}
}
\email{yuan.feng@unitn.it}

\author{Fabio Massacci}
\orcid{0000-0002-1091-8486}
\affiliation{%
  \institution{University of Trento}
  \city{Trento}
  \country{Italy}
}
\affiliation{%
  \institution{Vrije Universiteit Amsterdam}
  \city{Amsterdam}
  \country{Netherlands}
}
\email{fabio.massacci@ieee.org}

\title[Today's Cat Is Tomorrow's Dog]{Today’s Cat Is Tomorrow’s Dog: Accounting for Time-Based Changes in the Labels of ML Vulnerability Detection Approaches}

\begin{document}

\onecolumn
  
\newenvironment{project}[1]
{\par
 \bigskip
 \begin{wrapfigure}{l}[0pt]{1in}
 \vspace{-5pt}
 \includegraphics[width=1in,clip,keepaspectratio]{#1}
 \vspace{-25pt}
 \end{wrapfigure}
 \footnotesize\noindent}
{\par}

\newenvironment{bio}[1]
{\par
 \bigskip
 \begin{wrapfigure}{l}[0pt]{0.5in}
 \includegraphics[width=0.5in,clip,keepaspectratio]{#1}
 \vspace{-25pt}
 \end{wrapfigure}
 \footnotesize\noindent}
{\par}

\newenvironment{bio2}[1]
{\par
 \bigskip
 \begin{wrapfigure}{l}[0pt]{0.5in}
 \vspace{-15pt}
 \includegraphics[width=0.5in,clip,keepaspectratio]{#1}
 \vspace{-25pt}
 \end{wrapfigure}
 \footnotesize\noindent}
{\par}

\newenvironment{bio3}[1]
{\par
 \bigskip
 \begin{wrapfigure}{l}[0pt]{0.5in}
 \vspace{-10pt}
 \includegraphics[width=0.5in,clip,keepaspectratio]{#1}
 \vspace{-25pt}
 \end{wrapfigure}
 \footnotesize\noindent}
{\par}

\begin{figure}
\centering
\begin{minipage}{0.25\columnwidth}
\centering
\includegraphics[width=\linewidth]{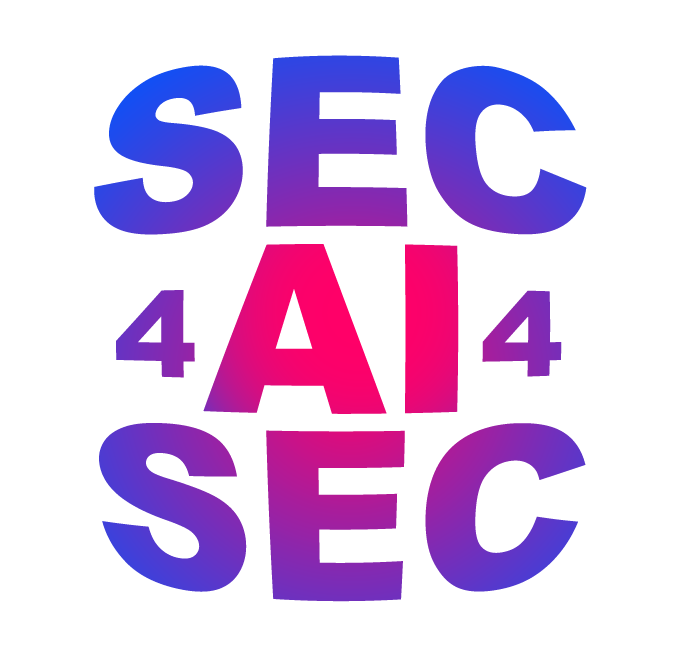}
\end{minipage}
\begin{minipage}{0.45\columnwidth}
\centering
\includegraphics[width=\linewidth]{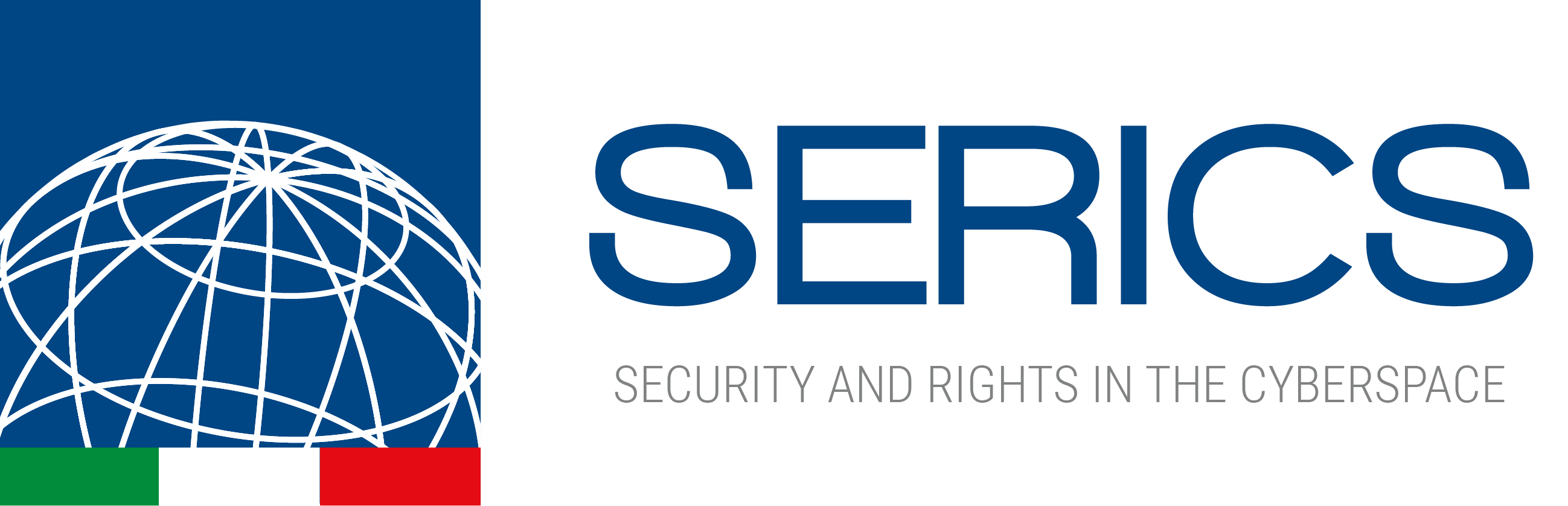}
\end{minipage}
\end{figure}

\vspace{2\baselineskip}

%\begin{figure}[h!]
%    \centering
%    \includegraphics[width=\textwidth]{logos/sec4ai4sec.png}
%\end{figure}

\vspace{2\baselineskip}

\begin{center}
{\huge \textbf{Today’s Cat Is Tomorrow’s Dog: Accounting for Time-Based Changes in the Labels of ML Vulnerability Detection Approaches}}
\end{center}

\vspace{\baselineskip}

{\large
Authors:
\begin{itemize}
    \item[]\textbf{Ranindya Paramitha}, University of Trento (Italy)
    \item[]\textbf{Yuan Feng}, University of Trento (Italy)
    \item[]\textbf{Fabio Massacci}, University of Trento (Italy), Vrije Universiteit Amsterdam (The Netherlands)
\end{itemize}
}

\vfill

\begin{figure}[h]
\vspace{-\baselineskip}
\includegraphics[height=3cm]{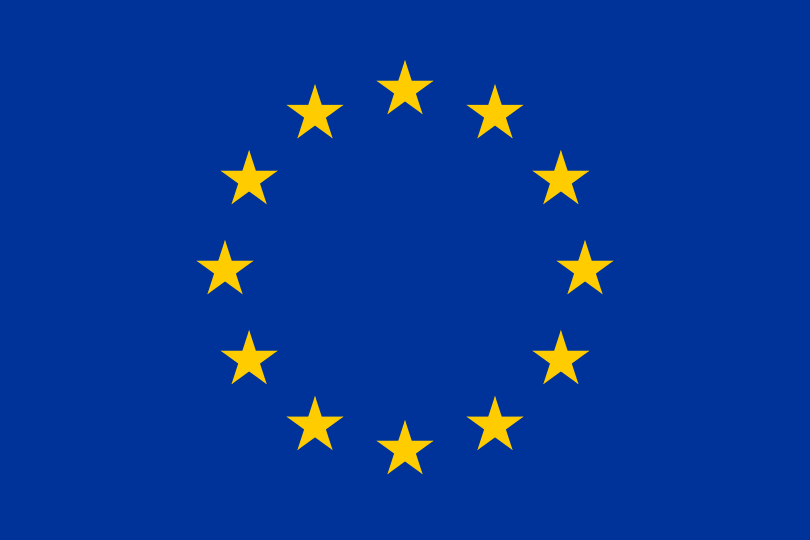}
~\includegraphics[height=3cm]{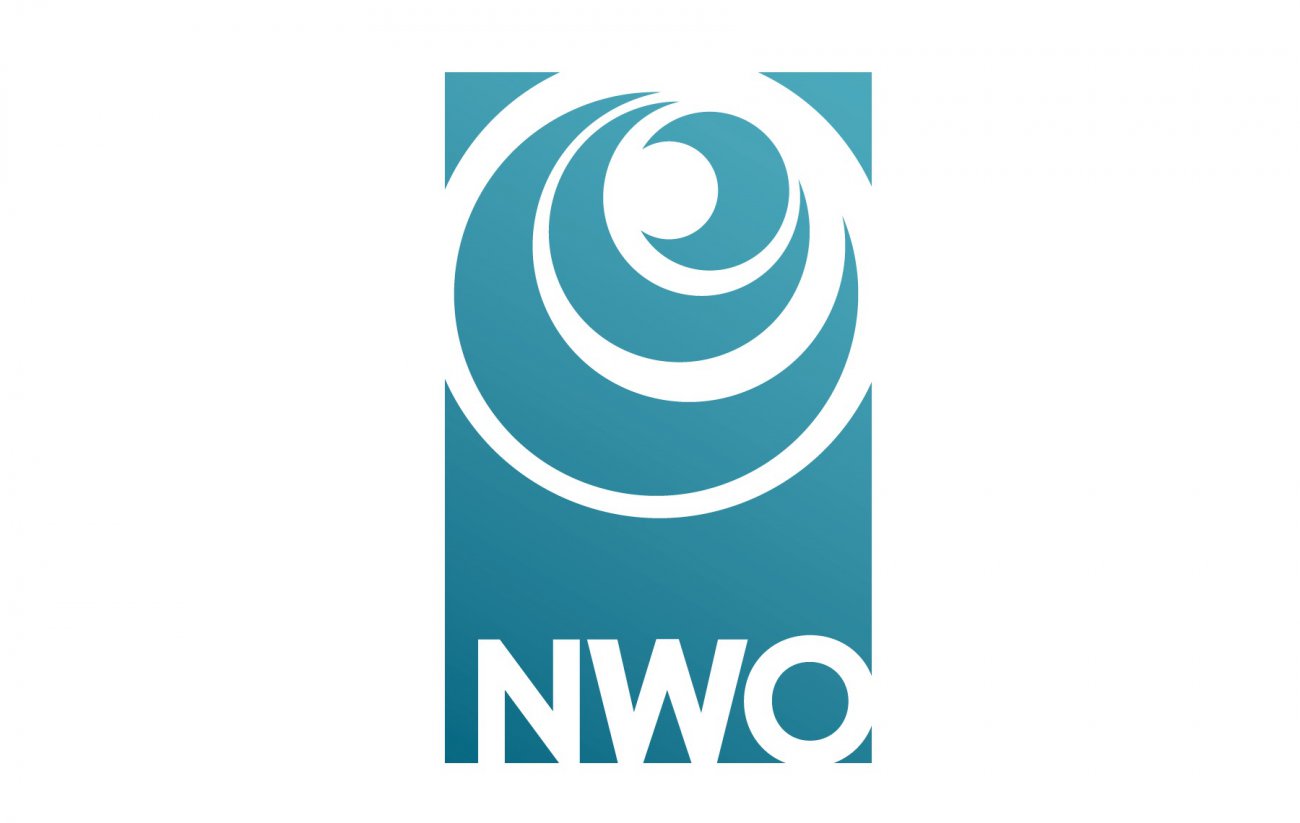}
~\includegraphics[height=3cm]{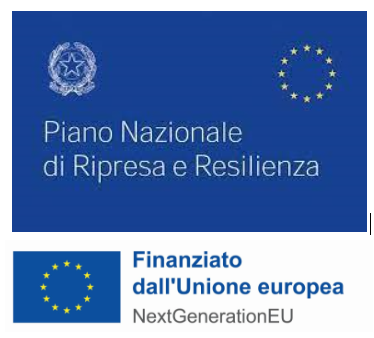}

\end{figure}

\noindent This work was partly funded by the EU under the H2020 Program AssureMOSS (Grant n. 952647) and the Horizon Europe Program Sec4AI4Sec (Grant n. 101120393), by the Italian Ministry of University and Research (MUR) under the P.N.R.R. – NextGenerationEU grant n.\ PE00000014 (SERICS subproject COVERT), and by the Dutch Research Council (NWO) under the grant NWA.1215.18.006 (Theseus) and grant KIC1.VE01.20.004 (HEWSTI). 
This paper reflects only the author's view, and the funders are not responsible for any use that may be made of the information contained therein.

\clearpage
\onecolumn

\begin{project}{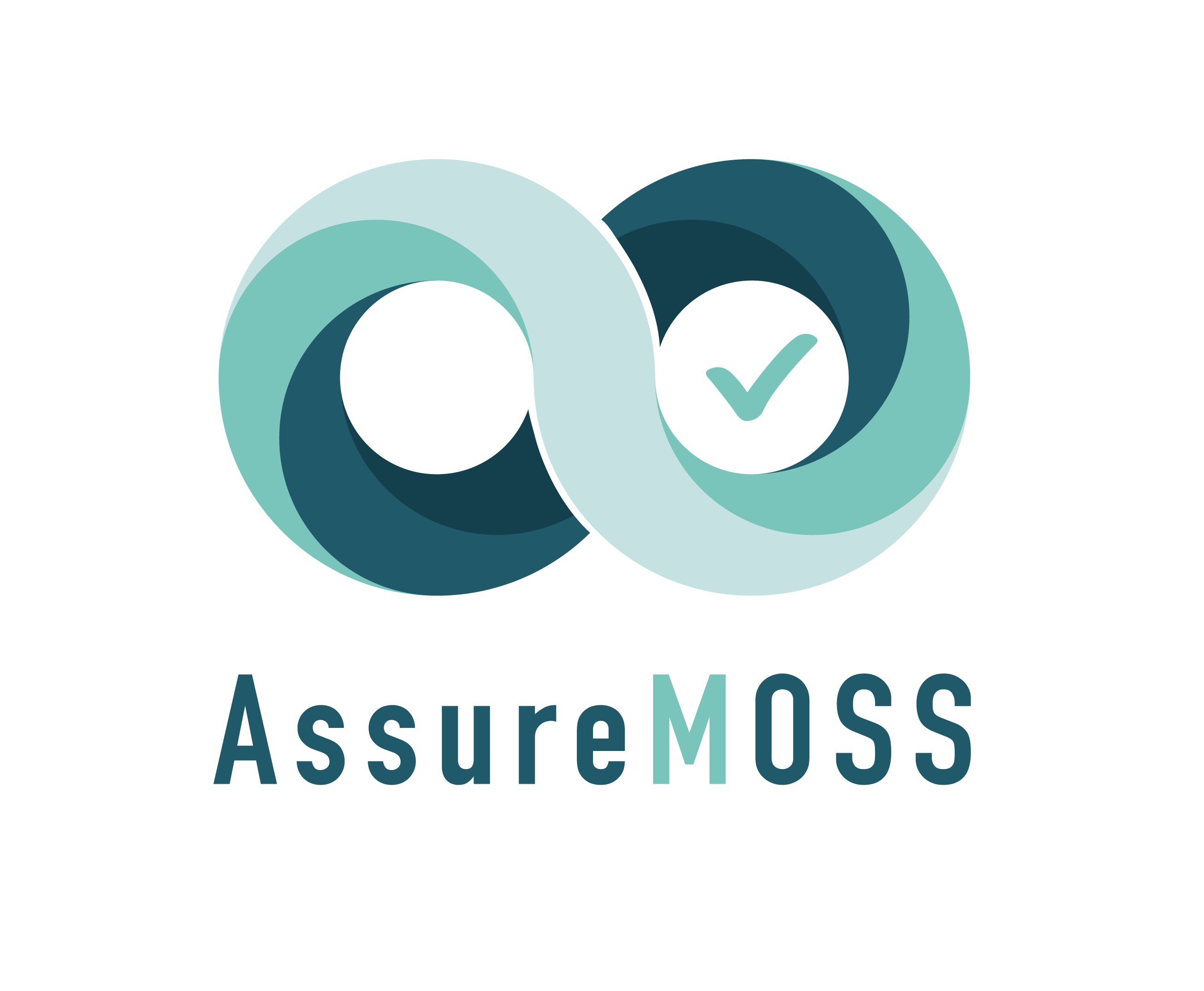}
    \textbf{Assurance and certification in secure Multiparty Open Software and Services (AssureMOSS)}. No single company masters its own national, in-house software. Software is mostly assembled from “the internet” and more than half comes from Open Source Software repositories (some in Europe, most elsewhere). Security \& privacy assurance, verification, and certification techniques designed for large, slow, and controlled updates, must now cope with small, continuous changes in weeks, happening in sub-components and decided by third-party developers one did not even know existed. AssureMOSS proposes to switch from process-based to artifact-based security evaluation by supporting all phases of the continuous software lifecycle (Design, Develop, Deploy, Evaluate, and back) and their artifacts (Models, Source code, Container images, Services). The key idea is to support mechanisms for lightweight and scalable screenings applicable automatically to the entire population of software components by Machine intelligent identification of security issues, Sound analysis and verification of changes, and Business insight by risk analysis and security evaluation. This approach supports the fast-paced development of better software with a new notion: continuous (re)certification. The project will generate also benchmark datasets with thousands of vulnerabilities. AssureMOSS: Open Source Software: Designed Everywhere, Secured in Europe. More information at \textbf{\url{https://assuremoss.eu}}.
\end{project}
% \vspace*{-6\baselineskip}
\begin{project}{logos/sec4ai4sec.png}
\textbf{Cybersecurity for AI-Augmented Systems (Sec4AI4Sec)} . As artificial intelligence (AI) becomes omnipresent, even integrated within secure software development, the safety of digital infrastructures requires new technologies and new methodologies, as highlighted in the EU Strategic Plan 2021-2024. To achieve this goal, the EU-funded Sec4AI4Sec project will develop advanced security-by-design testing and assurance techniques tailored for AI-augmented systems. These systems can democratize security expertise, enabling intelligent, automated secure coding and testing while simultaneously lowering development costs and improving software quality. However, they also introduce unique security challenges, particularly concerning fairness and explainability. Sec4AI4Sec is at the forefront of the move to tackle these challenges with a comprehensive approach, embodying the vision of better security for AI and better AI for security. More information at \textbf{\url{https://sec4ai4sec.eu}}.
\end{project}

\begin{project}{logos/Logo_Serics.png}
\textbf{In searCh Of eVidence of stEalth cybeR Threats
(COVERT)} aims to analyze emerging attack methodologies and develop advanced methods for detecting attacks and identifying guidelines for designing IT systems that ensure reduced vulnerability to new attack categories. The detailed objectives can be divided into four macro categories: (i) Development of advanced tools for analyzing malware and software aimed at identifying vulnerabilities that could be exploited by malware; (ii) Development of tools for analyzing network traffic to identify communications related to ongoing attacks; (iii) Development of machine learning systems that are robust to attacks and through which it is possible to extract knowledge aimed at creating more advanced tools for timely analysis and early identification of attacks; (iv) Analysis of the "human factors" involved in an attack with the development of tools for analyzing and correlating information from OSINT (open sources intelligence) and for the defense and prevention of attacks based on social engineering techniques.
\end{project}

\begin{project}{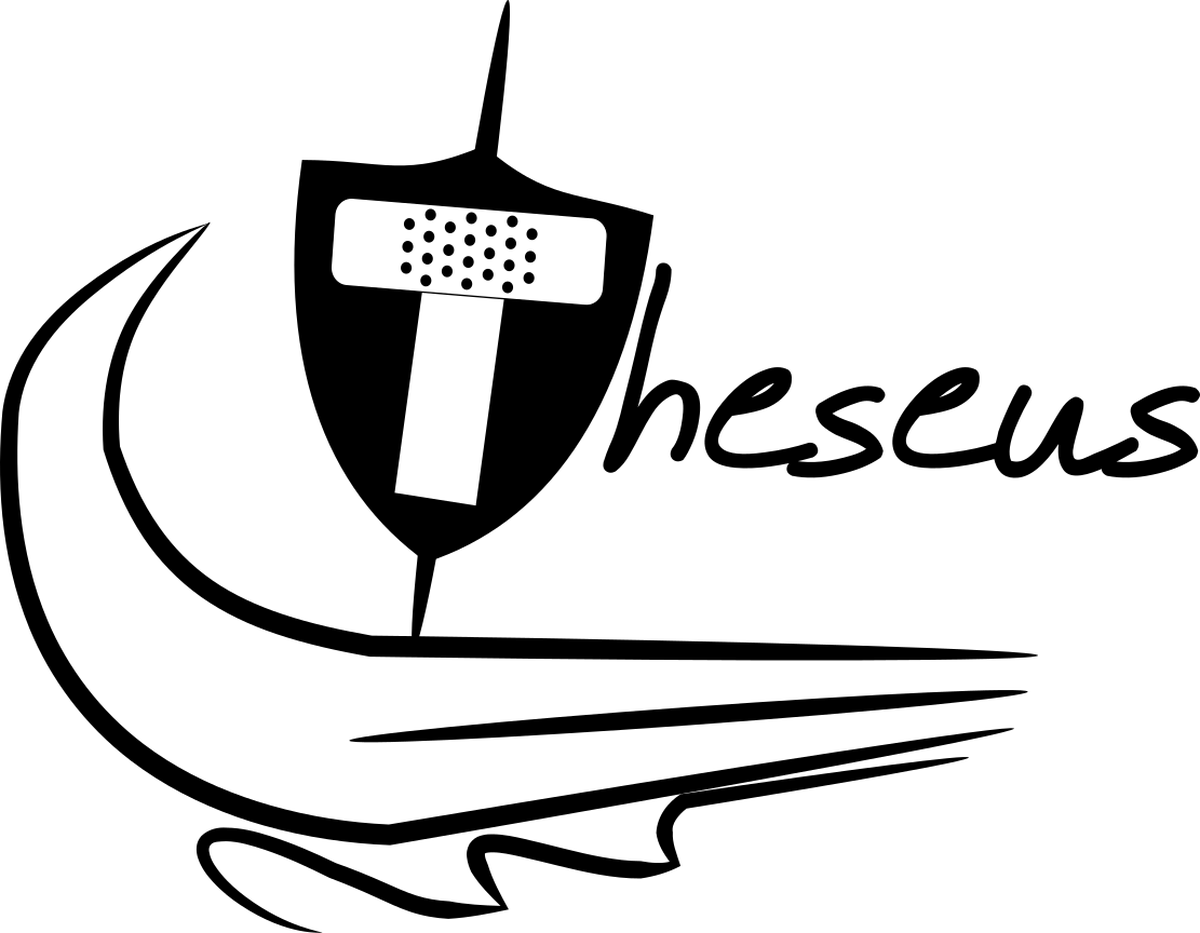}
    \textbf{Theseus: Making patching happen} project explores ways to improve software patching. The EU cybersecurity framework has evolved into a stricter, harmonized system that imposes requirements on patching. THESEUS has also found that tools like Shodan often misses vulnerabilities. Regulatory deadlines accelerate patching, although coordination issues cause delays. Researchers are developing efficient techniques to detect and mitigate vulnerabilities without performance loss. The reliability of patch-testing tools has proven to be a weak point, requiring improvements to reduce risks. Machine learning is being used to prioritize vulnerabilities more effectively. These insights contribute to better patching by addressing challenges in governance, technology, and operations.
\end{project}

\begin{project}{logos/NWO-logo.jpg}
\textbf{Hybrid Explainable Workflows for Security and Threat Intelligence (HEWSTI)}. In research into threats to safety and security, people and AI collaborate to obtain actionable intelligence. Their sources and methods often have significant uncertainties and biases. Experts are aware of these limitations, but lack the formal means to handle these uncertainties in their daily work. This project will invent a ‘metadata of uncertainty’ for threat intelligence (in both machine-readable and also human-interpretable forms) and validate it empirically. Intelligence agencies will then be able to explicitly consider the trade-off between the accuracy, proportionality, privacy, and cost-effectiveness of investigations. This will contribute towards the responsible use of AI to create a safer, more secure society.
\end{project}

\vfill

\begin{bio}{photos/paramitha}
\textbf{Ranindya Paramitha} (PhD 2025) is a research fellow at the University of Trento, Italy. She received her PhD from the University of Trento, Italy, in April 2025. Her main research interest is in software security, focusing on empirical analysis of secure software ecosystems, mining software repositories, and how developers can apply security. She is involved in a Horizon Europe Sec4AI4Sec project and has also started to actively serve the research community in several IEEE/ACM International Conferences/Workshops, such as by being a junior PC member (Distinguished Junior PC Award MSR'25) and regular PC member (ICSME’25). Contact her at \emph{ranindya.paramitha@unitn.it}.
\end{bio}

\begin{bio2}{photos/feng}
\textbf{Yuan Feng} is currently a PhD student at the University of Trento, Italy. Her research interests focus on Machine Learning for Vulnerability Detection (ML4VD) and security in software ecosystems. She is also interested in data quality issues within these domains. 
Contact her at \emph{yuan.feng@unitn.it}.
\end{bio2}

\vspace{\baselineskip}

\begin{bio3}{photos/massacci}
\textbf{Fabio Massacci} (PhD 1997) is a professor at the University of Trento, Italy, and Vrije Universiteit Amsterdam, The Netherlands. His research interests include empirical methods for the cybersecurity of sociotechnical systems. For his work on security and trust in sociotechnical systems, he received the Ten-Year Most Influential Paper Award at the 2015 IEEE International Requirements Engineering Conference. He is named co-author of CVSS v4. He leads the Horizon Europe Sec4AI4Sec project and the Dutch National Project HEWSTI.  He is the past chair of the Security and Defense Group of the Society for Risk Analysis, and an IEEE CertifAIEd Lead Assessor. Contact him at \emph{fabio.massacci@ieee.org}.
\end{bio3}

\vspace{\baselineskip}

How to cite this paper:
\begin{itemize}
    \item Paramitha, R., Feng, Y., and Massacci, F. Today’s Cat Is Tomorrow’s Dog: Accounting for Time-Based Changes in the Labels of ML Vulnerability Detection Approaches. \emph{Proceedings of the ACM on Software Engineering (PACMSE), Issue FSE 2025}. ACM Sigsoft.
\end{itemize}

License:
\begin{itemize}
\item This article is made available with a perpetual, non-exclusive, non-commercial license to distribute.
%\item The graphical abstract is an artwork by Anna Formilan.
\end{itemize}

\null\newpage

\begin{abstract}
% \textbf{[Context:]}
Vulnerability datasets used for ML testing implicitly contain retrospective information. When tested on the field, one can only use the labels available at the time of training and testing (e.g. seen and assumed negatives). As vulnerabilities are discovered across calendar time, labels change and past performance is not necessarily aligned with future performance. Past works only considered the slices of the whole history (e.g. DiverseVUl) or individual differences between releases (e.g. Jimenez et al. ESEC/FSE 2019). 
% \textbf{[Method:]}
Such approaches are either too optimistic in training (e.g. the whole history) or too conservative (e.g. consecutive releases). We propose a method to restructure a dataset into a series of datasets in which both training and testing labels change to account for the knowledge available at the time. If the model is actually learning, it should improve its performance over time as more data becomes available and data becomes more stable, an effect that can be checked with the Mann-Kendall test.
% \textbf{[Validation:]}
We validate our methodology for vulnerability detection with 4 time-based datasets (3 projects from BigVul dataset + Vuldeepecker’s NVD) and 5  ML models (Code2Vec, CodeBERT, LineVul, ReGVD, and Vuldeepecker). In contrast to the intuitive expectation (more retrospective information, better performance), the trend results show that performance changes inconsistently across the years, showing that most models are not learning.

\end{abstract}




\maketitle

\section{Introduction}
\label{sec:introduction}
Using machine learning for (security) bug detection is a recent popular trend in software engineering research~\cite{chakraborty2021deep,marjanov2022machine}. Despite the vigorous research, practical deployment is lagging~\cite{msr_keynote,lwakatare2020large,arp2022and}. For example, while several open source or commercial tools exist for static security analysis exist (e.g. SonarQuBE, Checkmarx), as well as open ML models for NLP (e.g.~\cite{liu2019roberta},~\cite{feng2020codebert},~\cite{hanif2022vulberta}), their transfer to security vulnerability prediction has not been uniformly successful ie. on the reproduction by ~\cite{steenhoek2023empirical}, CodeBERT~\cite{feng2020codebert} and VulbertA~\cite{hanif2022vulberta} have accuracy lower than 70\%. DiverseVul \cite{chen2023diversevul} reports a dismal F1 of 0.48 for the best LLM Model trained on a large dataset.

% \fm{add some cite of BERT models for security and papers showing Javabert model X is not working and is overfitting. Add a sentence if needed.}.

One possible explanation is that models are overfitting on different datasets~\cite{chakraborty2021deep,chen2023diversevul} or there are mistakes in the division of training and testing datasets~\cite{arp2022and,chen2019using}. However, Jimenez and colleagues~\cite{jimenez2019importance,jimenez2018engineering} have argued that there is a more fundamental cause that \emph{systematically} creates a gap.
Vulnerabilities and bugs have a published date, security-related commits have a commit date, etc. \emph{Before} the date on which the vulnerability was discovered (typically by somebody other than the developer) the `vulnerable fragment' was \emph{not known to be} vulnerable. The fix, which the ML algorithm will use as an example of not-vulnerable code, \emph{did not even exist}. 

Most high-quality datasets can be augmented with the date when the vulnerability was discovered, either by manual labeling (e.g., Devign~\cite{zhou2019devign}) or by using other publicly accessible data (e.g., CVEs from NVD~\cite{nvd}) as the ground truth of the labels. For example, BigVul~\cite{fan2020ac} uses vulnerability-fixing commits from real-life projects to identify a non-vulnerable fragment (the new version) and a vulnerable fragment (the version before the commit). 

Paradoxically, the major problem of all these datasets, whether synthetically created or based on real-life projects, is precisely that they provide at once the complete knowledge of vulnerable and non-vulnerable components at the time of investigation. Therefore, when a dataset is used with ten-fold cross-validation, 80-10-10 split, or other splits, \emph{the ML algorithm benefits from (retrospective) complete information}. The eventually correct labels of some fragments, that in field deployment of the ML model would have only been known in the future (sometimes after years \cite{ozment2006milk,massacci2011after,nguyen2016automatic,hu2024empirical}), are informing the training.

To address this issue, several studies~\cite{Shin_eval,Scandariato,jimenez2019importance} proposed to perform vulnerability prediction based on code metrics, keywords occurrences, function names, and other metrics (but not source code yet) and introduced the idea of restricting the predictor's knowledge to each release: training on the data from release 1 until $r$ and testing on $r+1$. Yet, they all tested using the labels from the complete information dataset, with the overall information across the years. This does not reflect the deployment scenario on a time $t$ where only a subset of the data and labels are actually known.
More recently, studies such as DiverseVul \cite{chen2023diversevul}, proposed to test ML models with an increasing share of the dataset, but they still sample the complete information dataset for training and testing. 

We need to extract from the complete information dataset
a slice that reflects the evolving knowledge that a model will encounter in time (new, unclassified yet, fragments, changing labels due to late discoveries). Hence, our first question:
\begin{description}
\item[\textbf{RQ1}:] \emph{How do we extract from a complete-information (retrospective) dataset a timeline of (perspective) datasets corresponding to what would be seen on the field at particular times?}
\end{description}
%\fi
We propose to use calendar time rather than releases (as in \cite{jimenez2019importance}) or random samples (as in \cite{chen2023diversevul}) for both correctness\footnote{Figure~\ref{fig:mot-exp} in our motivating example (\S\ref{sec:motivating_example}) shows how one release can have different labels in time, and therefore releases labeled with complete information would use future information. Further, some projects have frequent releases, sometimes even on the same day, and thus no learning could have reasonably taken place (except for fixing releases).} and causality\footnote{Releases and vulnerability discovery are independent processes collapsing on the times of responsible disclosure. By construction, the vulnerability timestamp in the NVD is \emph{after} (or on the same date) than the time of the new release. The fixing release is not an independent event from the published vulnerability discovery time.} considerations to form a timeline in which labels and data points for training and testing are different from the complete information source dataset and capture the partial information available at each time point. We propose to use calendar time instead of releases, also accounting for causality considerations. We train on vulnerability data from time $t_0$ until observation date $t$ and test on vulnerability data with labels available between the current and the next observation point $t+1$.  \emph{For each time point} in the investigation timeline we generate one training dataset and 2 testing datasets: 
\begin{enumerate}[leftmargin=*]
    \item[(1)] \Ideal all available code fragments up to the time of investigation are used for training, testing uses the labels known at the time of investigation. 
    \item[(2)] \Real test using vulnerable and non-vulnerable code and labels based on the information available in the next time period after the training. 
\end{enumerate}

So \IdealShort is the performance that researchers would obtain and report on papers by using the dataset available at the time of the investigation. 
On the other hand, \RealShort is the performance that one would observe by deploying the trained model and testing it on the field after the next time point observation period (e.g., one month, one year). 

Given our partitions, a consistent increase in performance (better precision and recall) is what one will expect from a technology to be used in practice as more and more data points are used in the learning process. The existence of such a trend can also be \emph{statistically tested} (with Mann-Kendall) thus showing that a model significantly learns over time. 

This finding would have been consistent with DiverseVul experiments \cite{chen2023diversevul}. Chen et al. Figure 3 shows that increasing the amount of training data from 10\% to 90\% of the sample improves the F1 score from 0.38 to 0.46. A key observation is that their procedure was a-temporal: the slices are randomly sampled from the \emph{same, complete-information retrospective dataset} and their labels do \emph{not} change from slice to slice (as they may do in reality). Hence, our second research question:
\begin{description}
\item[\textbf{RQ2}:] \emph{Does ML performance monotonically increase as the available (retrospective) information increases and consolidates?}
\end{description}
We implemented our proposed methodology for the specific case of a vulnerability dataset integrating 4 CVE-based datasets (Linux Kernel, OpenSSL, and Poppler from BigVul~\cite{fan2020ac} and
NVD dataset by Vuldeepecker~\cite{li_vuldeepecker_2018}) and a timeline date as input. 
Given a time-based dataset, we train 5 sota ML-based tools of different types (LineVul~\cite{fu2022linevul}, CodeBERT~\cite{feng2020codebert}, ReGVD~\cite{nguyen2022regvd}, Code2Vec~\cite{alon2019code2vec},  Vuldeepecker~\cite{li_vuldeepecker_2018}) and tested on vulnerability data using calendar years as period\footnote{Shorter intervals did not have enough changes and therefore the models would have essentially stayed identical.}.

In sharp contrast to the expectations, our analysis over the years shows that there is no significant trend of ML performance in detecting the vulnerability when tested in the \Real, which is the one people will experience at a certain point in time. We also show that presenting the ML performance result as (statistically tested) trends provides a more understandable visualization of the ML performance if deployed across the years.

\section{Motivating Example}
\label{sec:motivating_example}
\begin{figure}[t]
    \includegraphics[width=\columnwidth]{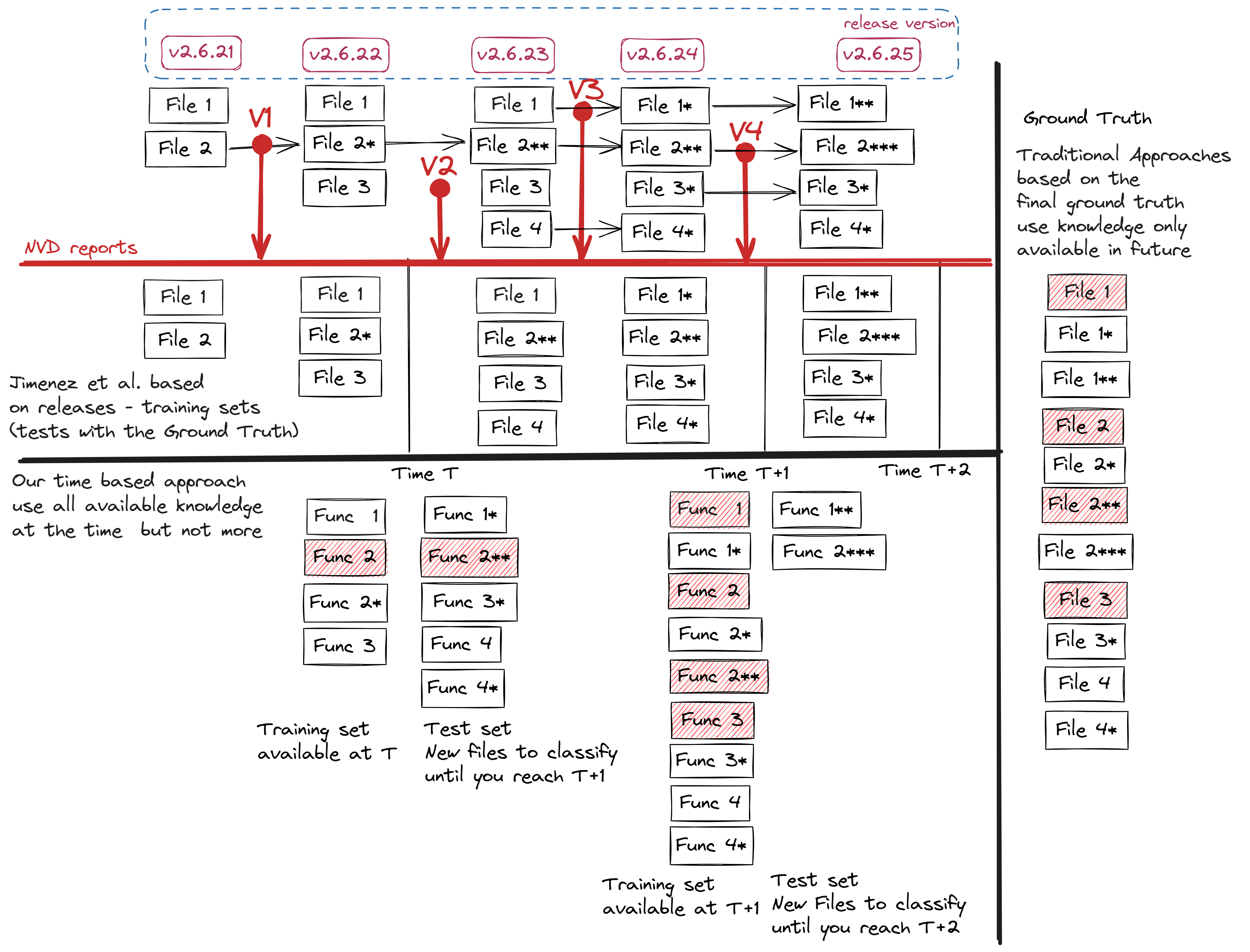}
    \longcaptionfig{\columnwidth}{
        
    }
    % \longcaptionfig{\columnwidth}{\small On mid-October 2023, \texttt{Linux Kernal} versions 4.0.9 and 4.0.10 were not known as vulnerable at that date: they were vulnerable in perspective. These cases should be labeled as `not vulnerable' in a dataset replicating the condition on which the algorithm will perform in the field. Otherwise, the ML algorithm will use information about something it should not know (yet).}
    \vspace*{-\baselineskip}
    \caption{Motivating example on the benefit of perspective: when you see what you should not see.}
    \label{fig:mot-exp}
    \vspace*{-\baselineskip}
\end{figure}

An example of how perspective will be implicitly present in a dataset and how this will change the performance of the ML algorithm is shown in Figure~\ref{fig:mot-exp}. 
We use five imaginary releases, inspired by the releases used by ~\cite{Shin_eval,Scandariato,jimenez2019importance}. These releases have some vulnerabilities reported in NVD, portrayed as red arrows in Figure~\ref{fig:mot-exp}. \cite{jimenez2019importance} did 2 experiments, ideal and real. The real experiment uses data from previous (one or three) releases in training, while the ideal experiment uses the complete information data. They claim their labeling (whether a file is vulnerable or not) is based on the reported vulnerabilities \emph{when a version is released}.
Therefore, when training on release v2.6.21, V1 would not be present in the training set (\textsc{File1} and \textsc{File2}, both are known as not vulnerable at that release). Also, in the testing set, \textsc{File3} will be considered as vulnerable as \cite{jimenez2019importance} always used complete information labeling to label their testing set. This labeling does not represent the scenario developers will face at that time, because at release v2.6.21 or v2.6.22, \textsc{File3} was still known as not vulnerable. Moreover, as a patch would most likely happen after a vulnerable version, if we see the training sets, the number of positive data points (vulnerable files) could be very low, for example, \textsc{File2} and its updates would not be considered as vulnerable in any training. For their ideal experiment, they always train and test using the data with complete information labeling.

Instead of using releases, we propose a methodology that uses \emph{time} as a method to split the dataset. We also use a different set to train and test in our experiments. We train on data and label available until time $T$ and then do 2 tests: retrospective test (using data and label available until time $T$) and perspective test (test with new data and label from $T$ to $T+1$). 

Our retrospective scenario represents training and testing using only available data until a certain time, while the perspective represents training using available data but testing on \textit{new} data and labels on the next time window. We are \emph{only} using the complete information data to test the last time point ($T_N$). Different timelines can produce different labels. This reflects what happens at a certain point in time, as vulnerability finding time will affect whether a vulnerability is known.

Additionally, we ran a third experiment with \emph{seen but believed negatives} data/label. We put the discussion on this additional experiment in a separate section (\S\ref{sec:ext}).

\section{Related Works}
\label{sec:related_works}
\subsection{Dataset for Detection ML Training}
\paragraph{Early works on bug/defect detection}
In their systematic literature review, ~\cite{pachouly2022systematic} mentioned that most of the early research on software defect prediction based on software metrics used NASA, PROMISE~\cite{promise}, and AEEM software defect datasets in their evaluation. The quality of such data was also disputed~\cite{shepperd2013data}.
Further, they do not use any issue/bug tracking system as their ground truth. A few datasets, such as Defects4J~\cite{defects4j} and iBugs~\cite{dallmeier2007extraction} use bug tracking systems in the version control system to find their label ground truth. According to~\cite{pachouly2022systematic}, also ~\cite{yatish2019mining} uses JIRA bug tracking systems as their ground truth. These datasets also provide information on when the bug is reported. More recent works \cite{Shin_eval,Scandariato,jimenez2019importance} have broadened the type and scope of datasets used.

\paragraph{Vulnerability detection}
\cite{chen2023diversevul} collected 10 vulnerability detection datasets.
Among the 10 datasets, SATE IV Juliet~\cite{okun2013report} and SARD~\cite{sard} are synthetic and semi-synthetic datasets that cannot fully represent real-world vulnerabilities, and they are used to evaluate several ML models~\cite{li_vuldeepecker_2018,mirsky2023vulchecker,russell2018automated}. 
On the other hand, Devign~\cite{zhou2019devign} is a manually labeled dataset consisting of 2 C projects (that are publicly available): FFMPeg and Qemu. Though a lack of a clear explanation of why a fragment is considered vulnerable, Devign has also been used by many ML models~\cite{zhou2019devign,feng2020codebert,hanif2022vulberta,ahmad2021unified,nguyen2022regvd}. 

To limit manual work, Draper~\cite{russell2018automated} and D2A~\cite{zheng2021d2a} utilize static analyzers to decide whether a code is vulnerable. However, the quality of these labels tends to be low (D2A's authors reported 53\% label accuracy. Other works utilizes security issues repository to generate a high-quality-labeled dataset such as \textsc{ReVeal}~\cite{chakraborty2021deep}, BigVul~\cite{fan2020ac}, CrossVul~\cite{nikitopoulos2021crossvul}, CVEfixes~\cite{bhandari2021cvefixes}, ProjectKB~\cite{ponta2019msr}, \textsc{DiverseVul}~\cite{chen2023diversevul}, and apart of Vuldeepecker's dataset~\cite{li_vuldeepecker_2018}. 
These datasets have higher accuracy labels, but  
model's evaluations based on them do not take into account the fact that they contain time-based information: some vulnerabilities are known (published) at a certain point in time. A summary of datasets and ML models is given in Table~\ref{tab:sota_ml-based}.

\begin{table*}[t]
    \centering
    \caption{Used datasets in the SOTA of ML-based vulnerability detection.}
    \longcaption{\textwidth}{Different ML models on vulnerability detection have been evaluated in the state of the art with different datasets. Some of them are time-based (labeling based on CVEs or issue trackers), and some are not. In our validation, we use the time-based dataset with CVE, and for our preliminary validation, we choose NVD, the smaller dataset. In bold are the dataset and models we use in our evaluation.}
    \resizebox{\textwidth}{!}{
    \begin{tabular}{p{0.28\textwidth}p{0.45\textwidth}p{0.09\textwidth}cp{0.37\textwidth}l}
        \hline
         & & & Has & \multicolumn{2}{c}{Used by}\\
          \multirow{-2}{*}{Dataset} & \multirow{-2}{*}{Description} & \multirow{-2}{*}{Labeling} & CVE? & ML & Architecture \\
         \hline\hline
         & & & & & \\
         & & CVE-based & & & \\
         \multirow{-4}{*}{\textbf{NVD~\cite{nvd}}} & \multirow{-4}{*}{\parbox{0.45\textwidth} {Fragments extracted from vulnerable software lined to the NVD Dataset managed by NIST.}} &  & \multirow{-4}{*}{\ding{51}} & \textbf{Vuldeepecker~\cite{li_vuldeepecker_2018}} & \textbf{RNN, BLSTM} \\ \hhline{----~~}
         & & Semi-synthetic & &SySeVr~\cite{li_sysevr_2022} & RNN \\
         \multirow{-3}{*}{SARD~\cite{sard}} & \multirow{-3}{*}{\parbox{0.45\textwidth} {Artificial dataset of test programs with documented weaknesses.}} &  & \multirow{-3}{*}{\ding{55}} &  & \\
         \hline\hline
         BigVul~\cite{fan2020ac} & A large C/C++ vulnerability dataset from FOSS Github projects. & CVE-based & \ding{51} & \textbf{LineVul \newline \cite{fu2022linevul}} & \textbf{Transformers} \\ \hline
         & & & & \textbf{CodeBERT~\cite{feng2020codebert}} & \textbf{Transformers} \\
         & & & & VulBERTa~\cite{hanif2022vulberta} & Transformers \\
         & & & & PLBART~\cite{ahmad2021unified} & Transformers \\\hhline{~~~~==}
         & & & & \textbf{Code2Vec~\cite{alon2019code2vec}} & \textbf{MLP, AST} \\ \hhline{~~~~==}
         & & & & \textbf{ReGVD~\cite{nguyen2022regvd}} & \textbf{GNN, token}\\
         & & & & Devign~\cite{zhou2019devign} & GNN, property graph  \\ \hhline{~~~~--}
         \multirow{-8}{*}{Devign~\cite{zhou2019devign}} & \multirow{-8}{*}{\parbox{0.45\textwidth} {Manually labeled dataset of vulnerable and non-vulnerable codes from 2 large-scale open-source C projects: FFMPeg and Qemu.}} & \multirow{-8}{*}{Manual} &\multirow{-8}{*}{\ding{55}} &  & \\ \hhline{----~~}
         Chromium+Debian \newline \cite{chakraborty2021deep} & A dataset of Chromium and Debian codes labeled by issue tracker. & Issue-based & \ding{55}  & \multirow{-1}{*}
         {\textsc{ReVeal}~\cite{chakraborty2021deep}} & \multirow{-1}{*}{GNN, property graph}\\ \hline
         DiverseVul \newline \cite{chen2023diversevul} & A C/C++ dataset from security issues websites, vulnerability-fixing commits, and source codes.& Issue-based & Some & Some LLM4Security such as LLMVulExp~\cite{mao2024towards} & LLM \\
         \hline
    \end{tabular}}
    \label{tab:sota_ml-based}
\end{table*}

\paragraph{Commit classification}
There are several manually labeled datasets on commit classification:
\cite{ghadhab2021augmenting} has created a manually labeled dataset of 1793 commits; \cite{dos2020commit} includes 3 manually classified datasets~\cite{levin2017boosting}, \cite{mauczka2015dataset}, \cite{safdari2018multiclass}. Other works tried to identify vulnerability-contributing commits (VCC) or security-relevant commits. \cite{perl2015vccfinder} by taking available CVEs with links to commits and tracing back the introducing commit (640 VCCs from 718 CVEs). A similar work is from~\cite{le2021deepcva}, which used NVD  to get 1,229 unique VCCs and manually curated some fixing commits (VulasDB~\cite{ponta2019msr}). 
% The difference is that they did a manual validation for a subset of samples and Le et al. managed to get twice as many (1,229 unique) VCCs. 
Moreover, TreeVUL~\cite{TreeVUL} categorized vulnerable types referring to the CWE tree and provided a fine-grained vulnerable type dataset. These datasets, which are based on an issue tracker, actually have a time component: the fix commits are only known when the vulnerability (with the fix) is published. Evaluating ML models using this kind of dataset makes it possible to exploit the time variable to avoid the benefit of hindsight.

\subsection{Models for Vulnerability Detection}

\paragraph{Learn Features}
\cite{li_vuldeepecker_2018} proposed Vuldeepecker which could find vulnerabilities that hadn't been reported in NVD. They built code gadgets and transformed them into vectors. The problem for Vuldeepecker is that it does not perform well in real-world situations~\cite{chakraborty2021deep}. And \cite{li_sysevr_2022} further proposed SySeVR to extract more related syntax and semantic vulnerability candidates automatically while reported in \cite{chakraborty2021deep}, the model's accuracy dropped by 73 percent when using real-world datasets.\cite{chakraborty2021deep} 
made a high-quality dataset with less duplication, more real data, and fewer irrelevant features in their model named \textsc{ReVeal}.

Other works are also engaged in learning more features. Devign~\cite{zhou2019devign} used the Conv module to help extract features for the GNN model in the model's classification process. 
Code2vec \cite{alon2019code2vec} used a code vector to predict semantic attributes. SvulD\cite{svuld} embedded subtle semantic information into the model which distinguishes and representative semantics are used. VulBG~\cite{vulbg} considers functions' features and behaviors to use in other functions' vulnerable detection. \cite{li2024effectiveness} considered context information. 

\paragraph{Text Classification}
After RoBERTa~\cite{liu2019roberta} and related pre-trained models significantly improved on natural language processing problems, CodeBERT~\cite{feng2020codebert}, VulBertA~\cite{hanif2022vulberta} and PLBert~\cite{ahmad2021unified} were proposed to work in source code's vulnerability detection. CodeBERT was the first to program language (PL) with natural language (NL) knowledge, which trained with a hybrid objective function. \cite{nguyen2022regvd} proposed ReGVD which used raw source code as tokens to build graphs. 
Moreover, LineVul~\cite{fu2022linevul} adopts 
Transformer techniques that can work on line-level predictions, other than on a function or file level. The results with DiverseVul \cite{chen2023diversevul} show that LLMs can only be competitive with GNNs for very large datasets. Interestingly, LLMs require a large amount of training data to
surpass ReVeal. When trained solely on CVEFixes data, a much
smaller training set, there is no clear advantage of LLMs over GNN-
based ReVeal model and ReVeal is even better than 6 LLMs (out of 10).

\iffalse
\paragraph{LLMs and GNNs} 
Chen et al.~\cite{chen2023diversevul} discussed that large language models(LLMs) perform better than existing Graph Neural Networks(GNNs) in vulnerability detection. In their observations, nowadays research on deep learning-based models does not prove that they can perform better than LLMs. Evaluated With a larger dataset, LLMs can achieve significant improvement over GNNs. They concluded that the influence of the amount of the dataset on the model's performance.  
\fi

\subsection{Model's Evaluation}

For the model's evaluation, previous works focused on two factors: metrics and operations \\ \cite{chakraborty2021deep}. Some other dimensions are considered based on the design of models. \cite{li2019comparative} proposed to do quantitative evaluations between different factors with two datasets. Besides, CodeBERT~\cite{feng2020codebert} was first evaluated by NL-PL tasks based on the model's specific structure. It was evaluated separately on both the NL side and the PL side. 

Moreover, Devign ~\cite{zhou2019devign} proposed baseline methods for the model's performance evaluation, including Metrics + Xgboost, 3-layer BiLSTM, 3-layer BiLSTM + Att, and CNN. This evaluation procedure was also adopted by ReGVD~\cite{nguyen2022regvd}. However, none of the previous evaluations actually take into account the time variable, which can exist in a vulnerability detection dataset as vulnerability is known at a certain point in time. 
% Therefore, the benefit of perspective was not addressed and the model training and evaluation on the past can be biased by the future.

\begin{table}[t]
    \centering
    \footnotesize 
    \caption{Comparison Between Previous Works}
    \label{tab:previousWork}
\begin{tabular}{p{0.1\linewidth}p{0.22\linewidth}p{0.18\linewidth}p{0.2\linewidth}p{0.18\linewidth}}
    \hline
    Study & Systems & Evaluation & Results & Labels  \\ \hline

    \cite{Shin_eval} & Mozilla Firefox, Red Hat Enterprise, Linux kernel & 
        Code metrics, Sample until each release &
        Precision .03 - .05, \newline Recall .87 - .90 \& .79 - .85
        & Until Release, \newline Complete Info\\          \hline

    \cite{Scandariato} & 20 Android apps & 
    Keywords, Sample until each release &
    Precision .90 \& .86, \newline Recall .77 
    & Until Release, \newline Complete Info\\          \hline

    \cite{Jimenez_2016} & Linux Kernel &
    Code metrics and function names/imports, Sample until each release &
    Precision .65 \& .76, \newline Recall .22-.64 \& .16-.48
    & Until Release, \newline Complete Info\\          \hline
    
    \cite{jimenez2019importance} & Linux Kernel, OpenSSL,  Wireshark &
    Code metrics and function names/imports, Sample until each release &
    Precision .45-.83, \newline Recall .36-.77
    & Until Release, \newline Complete Info\\          \hline

    \cite{chen2023diversevul} & {DiverseVul, Devign, ReVeal, BigVul, CrossVul, CVEFixes} &
    Source code functions, Random sample of complete info &
    F1 .09-.47\newline Precision .12-.52, \newline Recall .05-.44
    & Complete Info\\          \hline

    This paper & Linux Kernel, OpenSSL, Poppler, NVD Vuldeep. &
    Source code functions, Sample until each time &
    Recall 0.0-1.0, \newline FPR 0.1-.91 
    & Retrospective (Until Time $t$) \& Perspective (Until Time $t+1$)\\          \hline

    \end{tabular}
    
\end{table}

In Table~\ref{tab:previousWork}, we list some previous studies that worked on the evaluation method: release-based validation and cross-validation \cite{Shin_eval,Scandariato,Jimenez_2016}. They used metrics and keywords extracted from files as the prediction. The most complete work is by Jimenez and colleagues \citeyear{jimenez2019importance} who considered all different code metrics used by previous work. evaluated the model with consideration of different versions of packages that previous works did not count into. 

The closest papers to our work are the studies by Jimenez et al~\cite{jimenez2019importance} and Chen et al.\cite{chen2023diversevul}, we list different parts of methodology in Table  \ref{tab:cp_label}. Firstly, we work on a time-based dataset considering the CVE published date while DiverseVUl sample the overall dataset so it may contain vulnerability information that will be available in the future, while Jimenez et al. work on the released date of the repository (which does not account for the close, induced release of the CVE entry). The major difference is that \cite{jimenez2019importance,chen2023diversevul} \emph{always} test on the complete information data, when training both with the ideal or (so-called) real datasets. These data and labeling do not represent what developers will face in the real-life scenario, as they are not available yet. In contrast, we use data and labels available \emph{at each timeline} to compare between our retrospective (testing using data and label available at time $t$) and perspective (testing on data and label available between $t$ and $t+1$).  After experimenting with our datasets, we can see clearly how model performance would be influenced by the perspective data. 

\begin{table}[t]
    \caption{Comparison between Jimenez et al.~\citeyear{jimenez2019importance}, Chen et al \citeyear{chen2023diversevul} and this Paper}
    \longcaption{\textwidth}{}
    \centering
    \footnotesize
    \begin{tabular}{lp{0.225\textwidth}p{0.225\textwidth}p{.30\textwidth}}
        \hline
         & \cite{chen2023diversevul} & \cite{jimenez2019importance} & This Paper \\ \hline
         \multicolumn{3}{l}{\textbf{Methodology}} \\ \hline
         Training Set & 
         Random SubSample  & Sample until Release & Sample until Time \\
         Training Labels & Complete Info & Known at Release & Known at Time \\ 
         %Comparison &
       %  Ideal vs. Ideal & Ideal vs. Real & \IdealDataset vs. \RealDataset \\
         Testing Set & Sample of Complete Info & Sample until Next Release & Sample until Next Time \\
         Testing Label & Complete Info & Complete Info & Known at Next Time \\
         Results & Result as a trend (line chart) & Global results (boxplots) & Result as a trend, (line charts) \\ \hline
         \multicolumn{3}{l}{\textbf{Evaluation}} \\ \hline
         Systems(Datasets) & DiverseVUl, Devign, ReVeal, BigVul, CrossVul, CVEFixes & Linux Kernel, OpenSSL, Wireshark & Linux, OpenSSL, Poppler (from BigVul~\cite{fan2020ac}), NVD Vuldeepecker \\ 
         Models & 1 GNN, 10 BERTs & Random Forests & 2 BERTs, 1 GNN, 1 BLSTM, 1 MLP \\ \hline
    \end{tabular}

    \label{tab:cp_label}
\end{table}
% As a differentiated replication, this paper has some major differences from the previous paper. *\AddNegsDataset is \IdealDataset + added negatives. Added negatives are additional data points from the next year but as negatives (not vulnerable). The idea behind this is that usually the codes in these data points are already known but the vulnerability is not known yet.

\highlight{black}{\textbf{Main Gap:} Several datasets for ML evaluation have a time variable, but the current evaluation of ML models in SOTA often fails to take this variable into account.}

\section{Methodology}
% \nn{TODO:reconstruct}
\label{sec:methodology}

\begin{figure}[t]
    \centering
    \includegraphics[width=0.6\columnwidth]{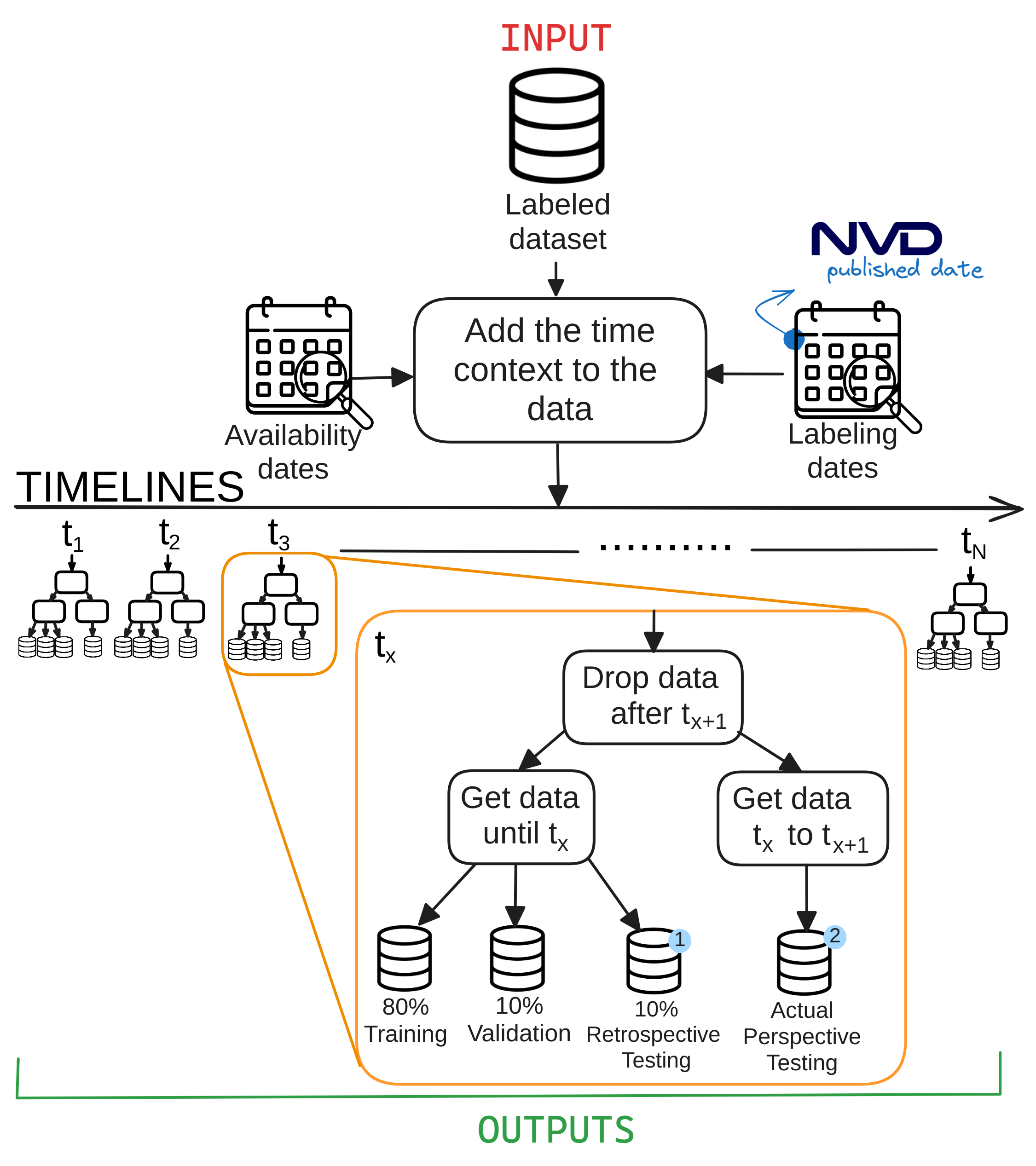}
\begin{minipage}[b]{0.3\linewidth}\footnotesize
{Given a time-based dataset, our proposed methodology adds time context to the data from a given source and then produces a timeline of datasets with perspective views. Time context can be instantiated with vulnerability publishing date, such as NVD, bug issue report date, etc. For each time point in the timeline, it produces one training set (possibly also one validation set) and 2 testing sets. It is important to note that we do not only split the data points by time, but also use only available labels at every point in time.}
\end{minipage}
    \caption{Our proposed methodology to eliminate retrospectives.}
    \label{fig:method}
\end{figure}

To answer RQ1, we propose a methodology to produce a timeline of datasets from a time-based labeled dataset, given a specific timeline of dates. The conceptual flow of the approach is depicted in Figure~\ref{fig:method}, and the inputs/outputs are summarized in Table~\ref{tab:input-output}.

\begin{table}[t]
    \centering
    \footnotesize
    \caption{Input and Output of Our Methodology}
    \begin{tabular}{p{0.17\columnwidth}p{0.47\columnwidth}p{0.26\columnwidth}}
        \hline
        Object & Description & Instantiation \\
        \hline
        \multicolumn{3}{l}{\textbf{Input}} \\
        \hline
        Labeled dataset & A dataset to train an ML-based vulnerability detection tool/ model with CVE information. These CVEs will then be used to determine whether the vulnerability is known or in perspective at the time of observation. & BigVul~\cite{fan2020ac}, Vuldeepecker's~\cite{li_vuldeepecker_2018} \\
        Availability dates & A set of dates that are pre-selected to simulate the information availability at a certain point in time for the training of the ML model & Published dates of the codes or commits. \\
        Labeling dates & A set of dates that are pre-selected to simulate the label availability at a certain point in time for the training of the ML model & Published dates of CVEs in the NVD. \\
        Timeline dates & A set of dates to simulate when we are observing the ecosystem. & 2010, 2011, 2012, ... \\
        Testing delta & (Optional)  a delta in months that we use to simulate the information availability at a testing time (which intuitively should be later than training time). & 6 months/ 12 months \\
        \hline
        \multicolumn{3}{l}{\textbf{Output} ($N$ time points)} \\
        \hline
        \multicolumn{3}{l}{For each time point in a timeline,} \\
        Training set & X\% of the relabeled dataset to train the ML model & 80\% + 10\% validation set\\
        2 testing sets $\times N$ & \IdealShort test set, \RealShort test set to test the ML model & 2 test sets $\times N$\\
        2 testing results $\times N$  & The result of testing the ML model on the 2 test sets $\times N$ & Precision, Recall (TPR), FPR, ... $\times$ 2 test sets $\times N$ \\
        \hline
    \end{tabular}
    
    \label{tab:input-output}
\end{table}

We elaborate on the steps of the methodology as follows (pseudocode is shown in Algorithm~\ref{alg:eliminate_hind}). 

\begin{enumerate}[leftmargin=*]
\item \emph{Add time context to data.} In this step, the data is mapped with the inputted labeling and availability dates. If the inputted labeling and availability dates come from another dataset, then the dataset is fetched before the mapping (e.g., a record labeled 1 now becomes labeled 0 at time $t$).
\item \emph{Produce a timeline of datasets to test ML's performance at each time point in a timeline.} We take a timeline date and the testing delta and use them as our reference. \textbf{For each time point in the timeline,} we produce a training dataset and a set of testing datasets from the input dataset based on different possible assumptions.
    \begin{description}[leftmargin=2ex,itemindent=-1ex]
        \item[\textbf{\Ideal}:] If the data point's labeling date is later than the chosen timeline, we drop positive and negative data points from the database. This rule assumes that before the timeline, the point does not exist.
        % \item[\textbf{\AddNegs}:] If a record's labeling date is later than the chosen timeline but the availability date is still inside the testing delta, we flip its label if it was previously labeled positive and drop it if it is labeled negative. We keep these records in the testing set. 
        \item[\textbf{\Real}:] We keep records with a labeling date later than the chosen timeline but still inside the testing delta. 
    \end{description}
\item \emph{\textbf{For each time point in the inputted timeline}, train the ML on the training set.} We train the ML on a subset of the relabeled dataset containing $X\%$ (predefined or provided as input) of data points from the complete info dataset. This subset can be divided into training and validation datasets.
\item \emph{\textbf{For each time point in the inputted timeline}, test the ML model on 2 different scenarios} mentioned in Step 2: \IdealShort and \RealShort.
\end{enumerate}

\begin{algorithm}[t]
\vspace{0.5\baselineskip}
\scriptsize
\SetKwInOut{Input}{input}\SetKwInOut{Output}{output}
\SetKwProg{Fn}{Function}{:}{end}
\Input{Labeled dataset ($original\_dataset$)}
\Input{Labeling dates ($label\_dates$)}
\Input{Availability dates ($availability\_dates$)}
\Input{Timeline dates ($timeline\_dates$)}
\Input{(Optional) Testing delta ($delta$ in months, default=12)}
\Input{(Optional) Percentage ($percentage$)}
\Output{A Timeline of datasets ($datasets$)}

\BlankLine
\tcp{Add time context to data.}
$dataset \leftarrow mapDataWithTime(original\_dataset, label\_dates, availability\_dates)$

\BlankLine
\tcp{The function to get 3 testing datasets.}
\Fn{$relabel(dataset, timeline\_date)$}{
    $\RealTestDataset\_test \leftarrow []$
    
    \For{$ rec | rec \in dataset$}{
        \uIf{$rec.label\_date > timeline\_date + delta$}{
            $dataset.dropData(rec)$
        }
        \uElseIf{$rec.label\_date > timeline\_date$}{
            \tcp{assumption == R-P}
            $\RealTestDataset\_test.append(rec)$
            
            \tcp{For Retrospective: Both label and record unknown: we drop the record}
            $dataset.dropData(rec)$ 
        }
    }
    \Return $dataset, \RealTestDataset\_test$
}

\BlankLine
\tcp{Produce a timeline of datasets.}

$datasets \leftarrow \{\}$

\For{$tdate|tdate \in timeline\_dates$}{
    $dataset, \RealTestDataset\_test \leftarrow relabel(dataset, tdate)$ 

    $datasets[tdate][``perspective"] \leftarrow \RealTestDataset\_test$
    
    \BlankLine
    \tcp{Split into training and testing.}
    $training,  validation, testing\_dataset \leftarrow splitTrainTest(dataset, percentage)$ 

    $datasets[tdate][``training"] \leftarrow training$
    
    $datasets[tdate][``validation"] \leftarrow validation$
    
    $datasets[tdate][``retrospective"] \leftarrow \IdealTestDataset\_test$
}
\Return $datasets$
\caption{Produce a timeline of datasets.}
\label{alg:eliminate_hind}
\end{algorithm}

We can then plot the metrics from the tested model, both for \IdealShort case and \RealShort case, to see if there is a certain trend or difference. For this comparison, any metrics would do because basic metrics (TP, FP, TN, and FN) are not independent: at least their sum must be equal to the total number of samples. In the general case, there are more complex dependencies due to the chosen ML algorithm. Therefore, if one aggregates these basic metrics to form other metrics (such as precision, recall, and F1) and one metric fluctuates, the other metrics would also fluctuate in balance.

\section{Dataset and Model Selection}
\label{sec:selection}

\subsection{Dataset}
\label{subsec:dataset}
\begin{figure}[t]
    \centering
\includegraphics[width=0.5\textwidth]{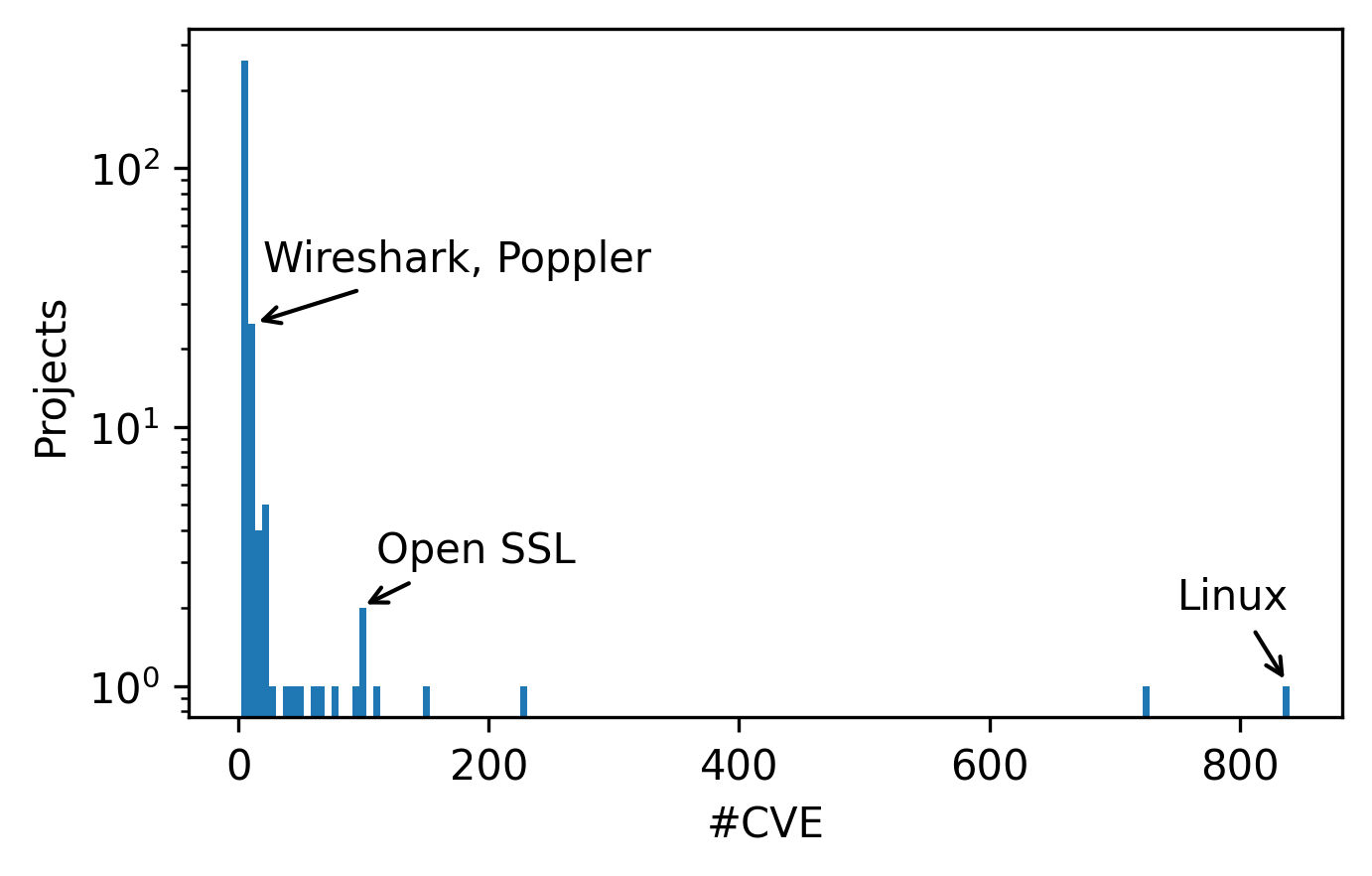}\hspace{4ex}
\begin{minipage}[b]{0.45\linewidth}\footnotesize
    {The 3 projects chosen by~\cite{jimenez2019importance} cover the ecosystem well, as \texttt{linux} can represent the projects with high number of CVE, \texttt{openSSL} for projects with medium number, and \texttt{wireshark} for the low number. As \texttt{wireshark}'s CVEs are published in the same year, we take another project from the same position \texttt{poppler} to represent the projects with a low number of CVEs}
    \vspace{8ex}
\end{minipage}
    \caption{Distribution of CVE count (\#CVE) in BigVul Dataset}
    \label{fig:bigvul-dist}
\end{figure}

We wanted to replicate the study by~\cite{jimenez2019importance} (the study with a similar idea to ours) using our methodology. However, their dataset does not include the publication date and/or CVE ID of the vulnerability, which makes our methodology inapplicable. We also tried to use VulDat7~\cite{jimenez2018engineering}, which is the framework used by~\cite{jimenez2019importance} to generate their dataset, but the framework is outdated and cannot be used. Therefore, we decided to instead use the BigVul dataset~\cite{fan2020ac}, which covers real vulnerabilities until 2019. Figure~\ref{fig:bigvul-dist} shows the distribution of the number of CVEs (\#CVE) by project. It also shows the position of the 3 projects used by~\cite{jimenez2019importance}: \texttt{linux}, \texttt{wireshark}, and \texttt{openSSL} in the distribution. 

In BigVul, \texttt{wireshark} has 10 CVEs, but all of them are published in the same year. Therefore, \texttt{wireshark} is not a good example to show how our methodology shows the performance difference between the realistic and the ideal world, as we do not have the data to test in the realistic setting (assuming we train on $y$ and test on $y+1$). We then decided to choose another project from the same position in the distribution (representing projects with low CVE count): \texttt{poppler}, whose number of CVEs is in the same group as \texttt{wireshark} but its CVEs are published in different years spanning from 2009 until 2018.

Additionally, we reviewed 10 different papers from the SOTA on ML-based vulnerability detection evaluated by \cite{steenhoek2023empirical}. We then clustered them based on the datasets they used in Table~\ref{tab:sota_ml-based}. This table has covered different ML architectures, ie. RNN, MLP, GNN, and transformers. We then decide to also evaluate our methodology on the NVD dataset by Vuldeepecker~\cite{li_vuldeepecker_2018} which can be found in Github~\cite{vdp-data} as it is lightweight and suitable for a preliminary evaluation. We chose this dataset as it uses real-world vulnerabilities (CVE-based) like BigVul dataset~\cite{fan2020ac}, and is not manually labeled like Devign's dataset~\cite{zhou2019devign}.
From the NVD dataset, we filtered to get only the records with CVEs. Vuldeepecker provided 2 datasets: CWE-119 and CWE-399, which we combined to get more records. The merged dataset contains 628 records with 68 CVEs in total.

\subsection{Models}
\label{subsec:models}
In Table~\ref{tab:sota_ml-based}, we reported a selection of historical works to illustrate the evolution of the field. Each line differs from the previous ones in either using a different representation or dataset. In recent years, we have seen a consolidation of some preferred methods (BERT and LLMs in general) and a stronger attention to the development of datasets \cite{risse2024uncovering}, which have evolved over the years. Despite the different representations adopted, the input code is often flattened into a vector when serving as input for a neural network. Models also evolved: after the first BERT-based detection model~\cite{hanif2022vulberta} or variants tested different languages~\cite{desousa2021javabert} used previous models as a pre-trained model~\cite{vdet}.

We then picked five ML models (Vuldeepecker, ReGVD, CodeBERT, Code2Vec and LineVul) with different ML architectures out of other models in Table \ref{tab:sota_ml-based} for our evaluation. We picked them to see how different ML architectures (BLSTM, GNN, MLP, and Transformers) react to the elimination of retrospectives that we did. 
We ran the replication package by~\cite{steenhoek2023empirical} for all models except Vuldeepecker, for which we ran an implementation in Python available on Github~\cite{vdpython}.

\section{Implementation}
\label{sec:implementation}
To see the methodology in action, we implement a Python script for a specific use case: \emph{vulnerability detection dataset}. 
Our Python implementation needs several inputs that are mapped to the inputs in our methodology:    
\begin{figure}[t]
    \centering
    \includegraphics[width=0.6\columnwidth]{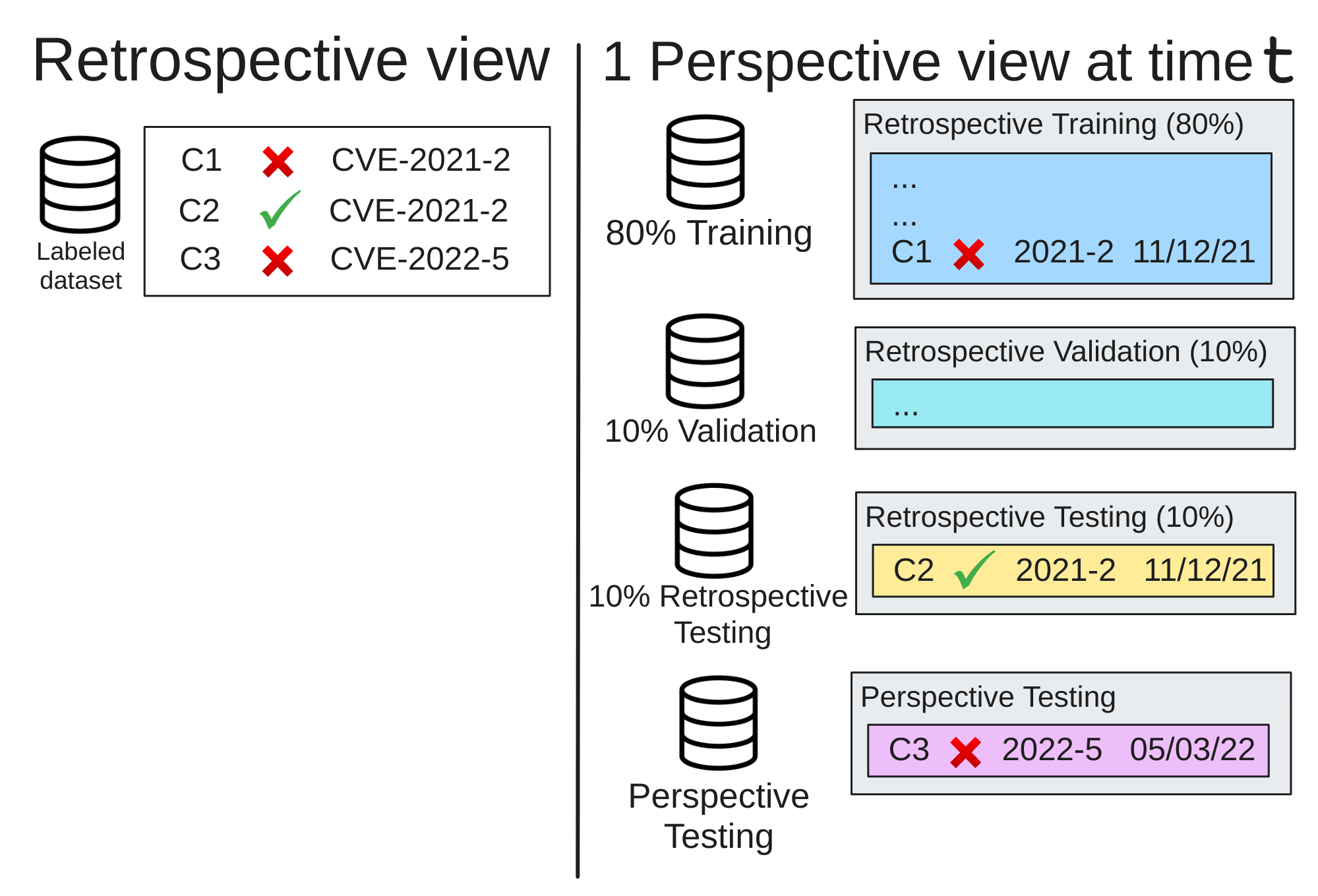}
\hspace{4ex}\begin{minipage}[b]{0.35\linewidth}
\footnotesize One record on a dataset is a code fragment with a CVE label. The code availability dates are not available, so we assume that the code is available before the code is labeled by a CVE (vulnerable or not vulnerable). The date on the record is the date when a CVE is published for a certain code. The negative records also have a reference to a CVE because they are extracted from the same commit as the vulnerable code mentioned in the CVE. From one input dataset, we produce a timeline of datasets, of which for each time point, we produce 1 training (and 1 validation) set and 2 testing sets. (\IdealTestDataset and \RealTestDataset).
\end{minipage}
    \caption{Example instantiation of the proposed methodology.}
    \label{fig:example}
\end{figure}
\begin{itemize}[leftmargin=*]
    \item \textbf{Labeled dataset}: refer to Subsection~\ref{subsec:dataset} of dataset selection.
    \item \textbf{Labeling dates}: for the code labeling dates (vulnerable/ not vulnerable), we use NVD~\cite{nvd}'s published date. For any vulnerability in the input dataset with an unknown published date, we fetch it directly through NVD's APIs~\cite{nvd-api}.
    \item \textbf{Availability dates}: as the original dataset we use does not have information about the code availability dates, we assume that the code availability is less than or equal to the labeling dates (NVD published date). This assumption holds as the code has to be available before a CVE can be found. For this implementation, we use labeling dates as availability dates. We acknowledge this as a possible threat to validity in Section~\ref{sec:tov}.
    \item \textbf{Testing delta:} we use 12 months as we assume that the machine learning trained in $y_0$ will be used (tested) to identify vulnerability in the next year ($y_1$).\footnote{A shorter timeline, e.g., every quarter or even every month, is an interesting option, but its prediction will oscillate too strongly as not many code fragments are committed in our dataset. A larger dataset with daily or monthly updates would be needed, and we plan to design it for future work.}
    \item \textbf{Split train-test percentage}: this input adds flexibility for the user of the script to define how to split the dataset into training and testing, and also possibly validation datasets. The default is 80\% training, 10\% validation, and 10\% testing.
\end{itemize}

The script is implemented with the following steps:
\begin{enumerate}[leftmargin=*]
    \item \textit{If the input dataset is not completely CVE-based}: Take only the CVE-based part of the dataset to be processed. 
    \item Remove unavailable information from the data and relabel: \textbf{for each time point in the timeline}, we use the assumption rules mentioned in Section~\ref{sec:methodology} to assess 3 kinds of performances. So for each record in the dataset
    \begin{itemize}
        \item[P1] \Ideal. We check on the NVD when the CVE is published. If the published date of the CVE is later than the timeline date, we drop the records. This simulates the training at a certain point in time by using only records with known labels at that time.
        \item[P2] \Real. We keep only records in the testing delta after the chosen timeline. This dataset simulates the real-life testing as a model that we trained on period $t$ would most likely be used in production to detect vulnerabilities during period $t+1$.
    \end{itemize}
    The example of the input retrospective dataset and one of the produced perspective dataset (for time $t$) is shown in Figure~\ref{fig:example}.
    \item Split train-test: based on the input, we split the datasets into training and testing (also validation if necessary). For the second assumption, we merge the testing dataset from the splitting with the added negatives.
    \item Return the outputs: 1 training set (can be divided into training and validation), and 2 different testing sets: \emph{\IdealTestDataset} and \emph{\RealTestDataset}.
    \item Train and test (2 times) the ML models.
\end{enumerate}

\section{Evaluation}
\label{sec:eval}

This section presents our methodology evaluation with the security vulnerability detection use case and how it answers our research questions.

\begin{table}[t]
    \centering
    \footnotesize
    \caption{Produced datasets for training and testing}
    \longcaption{\columnwidth}{The datasets grow more and more because the code availability and the CVEs found are also getting more and more throughout the years. However, the proportion of positive and negative data points remains stable.}
    \resizebox{\columnwidth}{!}{
    \begin{tabular}{l|rrrrrrrrrrrrr}
    \hline
    Dataset & Metric & 2008 & 2009 & 2010 & 2011 & 2012 & 2013 & 2014 & 2015 & 2016 & 2017 & 2018 & 2019\\ 
        \hline
        & Training & -& - & -  & 9&  7668 & 13274 & 17234&  20871 & 27548&  32159 & 35315 & 38014    \\
        & Validation & -&  - & -  &  2    &    853       &  1476     &    1916    &     2319 &        3062      &   3574 &        3925      &   4225   \\ 
        & Retrospective Testing& -& - & -  &  2    &     948    &     1640    &     2128    &     2578 &        3402  &       3971  &       4361     &    4694  \\ 
        \multirow{-4}{*}{\texttt{Linux}} & Perspective Testing& -& - & -  & 9456    &    6921    &     4888  &       4490       &  8244    &     5692  &       3897      &   3332   & -      \\
        \hline

        & Training &-&  -   &     -   &     -   &     -     &   28     &   247   &     780     &   1163    &    1308    &    1340      &  1505\\ 
        & Validation  &-&  -    &    -    &    -     &   -    &    4       & 28    &    88   &     131   &     147    &    149   &     168    \\ 
        & Retrospective Testing&-&  -     &   -       & - &        -   &     4     &   31    &    98     &   145     &   163  &      166    &    187  \\ 
       \multirow{-4}{*}{\texttt{OpenSSL}} & Perspective Testing  &-&  -      &  -   &     -   &     -    &    270  &      660   &     473       & 179     &   37    &    205   &     - \\ \hline

        & Training & -& 475  &       578     &    578   &      578    &     814    &     844   &      844      &   844 &        872     &    873 & - \\ 
        & Validating  &-&  53   &      66    &     66      &   66    &    91      &   95    &     95      &   95      &   98 &        98  & - \\ 
        & Retrospective Testing &-&  60  &       73    &     73      &   73    &     102    &     105  &       105  &       105     &    109   &      109  &  - \\ 
       \multirow{-4}{*}{\texttt{Poppler}} & Perspective Testing &-&  129   &      -      &   -     &    290     &    37   &      -     &    -     &    35    &     1     &    - & -\\ \hline

        & Training & 146    &    249    &    300   &     416   &     663     &   941     &   1175   &     1691     &   2163     &   2192   & -&     - \\ 
        & Validating  & 18   &     31    &   37  &      52    &    82       & 118       & 146    &    211     &   271      &  274    & -&    -    \\ 
        &Retrospective Testing & 19   &     32    &    39    &    53   &     84 &       118    &    148      &  212   &     271  &      276 &  -&      -     \\ 
       \multirow{-4}{*}{\texttt{NVD Vuldeep.}} & Perspective Testing  & 129  &      64      &  145     &   308  &      348    &    292     &  645      &  591    &    37   &    -   &  -&    - \\ \hline
        
    \end{tabular}}
    \label{tab:dataset_with_amount}
\end{table}

\begin{table}[t]
    \centering
    \footnotesize
    \caption{Dataset generation across the years}
    \longcaption{\columnwidth}{The mean derivative shows the increase of data points every year, which reaches 48.6\%. The \textit{Never seen} columns show how the test in the \RealTestDataset\xspace dataset contains way more data points compared to the \IdealTestDataset\xspace dataset (\textit{Seen and known} columns).}
    \footnotesize
    \resizebox{\columnwidth}{!}{
        \begin{tabular}{l|p{0.09\columnwidth}|
        >{\raggedleft\arraybackslash}p{0.13\columnwidth}
        >{\raggedleft\arraybackslash}p{0.13\columnwidth}
        >{\raggedleft\arraybackslash}p{0.13\columnwidth}|
        >{\raggedleft\arraybackslash}p{0.13\columnwidth}|
        >{\raggedleft\arraybackslash}p{0.11\columnwidth}
        >{\raggedleft\arraybackslash}p{0.11\columnwidth}}
        \hline
            & & \multicolumn{3}{c|}{Fraction with Vulnerabilities} & & & \\
            Dataset & Metric &  \multicolumn{1}{l}{\parbox{0.13\columnwidth}{Seen and known Positive Training $t$}} &	 \multicolumn{1}{l}{\parbox{0.13\columnwidth}{Seen and known Positive Validation $t$}} &	 \multicolumn{1}{l|}{\parbox{0.13\columnwidth}{Seen and known Positive Present Test $t$}} &  \multicolumn{1}{l|}{\parbox{0.13
            \columnwidth}{Seen but Believed Negative at that time}} & \multirow{-2.2}{*}{\parbox{0.11\columnwidth}{Never seen Positive, tested $t+1$}} & \multirow{-2.2}{*}{\parbox{0.11\columnwidth}{Never seen Negative, tested $t+1$}} \\ \hline
            
            & Mean Drv. & 28.9\% & 28.7\% & 28.4\% & 75.9\% & 75.9\% & 49.2\%\\  \hhline{~-------}
            & \multicolumn{7}{c}{Relative \%} \\
            \hhline{~-------}
            & Mean & 3.2\% & 0.4\% & 0.4\% & 1.1\%& 4.2\% & 95.8\%\\
            \multirow{-4}{*}{\texttt{Linux}} & St.Dev & 0.2\% & 0.0\% & 0.0\% & 0.9\% & 0.9\% & 0.9\%\\ 
            
            \hline

            & Mean Drv & 50.3\%  & 48.1\%  & 54.5\% & 31.8\% & 31.8\%  & 64.9\% \\  \hhline{~-------}
            & \multicolumn{7}{c}{Relative \%} \\ \hhline{~-------}
            & Mean & 8.2\%  & 1.1\%  & 1.2\%  & 33.5\%  & 10.6\%  & 89.4\% \\
            \multirow{-4}{*}{\texttt{OpenSSL}} &St.Dev & 2.6\%  & 0.7\%  & 0.7\%  & 33.4\%  & 6.1\%  & 6.1\% \\ 
            
            \hline
            
            & Mean Drv & 7.3\%  & 2.8\%  & 6.5\% & 27.8\% & 27.8\%  & 31.7\% \\  \hhline{~-------}
            & \multicolumn{7}{c}{Relative \%} \\   \hhline{~-------}
            & Mean &  2.8\%  & 0.4\%  & 0.4\% & 0.3\% & 0.3\%  & 52.8\% \\
            \multirow{-4}{*}{\texttt{Poppler}} &St.Dev & 0.2\%  & 0.0\%  & 0.0\% & 0.5\% & 0.5\%  & 50.2\% \\ 
            
            \hline
            
            & Mean Drv & 38.5\%  & 39.5\%  & 37.6\% & 92.6\%& 92.6\%  & 81.5\% \\\hhline{~-------}
            & \multicolumn{7}{c}{Relative \%} \\\hhline{~-------}
            & Mean & 27.3\%  & 3.3\%  & 3.5\%  & 14.2\%& 14.2\%  & 60.3\% \\
            \multirow{-4}{*}{\texttt{NVD Vuldeep.}} & St.Dev & 4.1\%  & 0.6\%  & 0.4\% &  8.2\%& 8.2\%  & 10.2\% \\
        \hline
        \end{tabular}}
    \label{tab:rq1}
\end{table}

\subsection{Dataset Generation}
\label{sec:rq1}

We first generated 2 datasets for each timeline and for each project using our Python script that implemented our methodology. The generated datasets are shown in Table~\ref{tab:dataset_with_amount}. 
From the generated datasets across the years, we can have several observations as portrayed in Table~\ref{tab:rq1}.

Even with the data point increase, the relative mean and standard deviation show that the percentage of positive (vulnerable) data points is more or less the same yearly.
The \textit{Never seen Positive/Negative, tested next year} columns have the data from the next year to simulate testing a previously trained ML model in the next year's data. 
On average, every year we test 6.2\%-14.9\% of the total population of negatives that have never been seen. The mean derivative of the yearly amount of data points shows that it increases 31.7\%-81.2\% every year. With complete information, we test with only 0.2\%-4.2\% of negatives every year, while if we consider the information available at any given time, we test with 52.8\%-95.8\% of the negatives that have never been seen.

\highlight{black}{\textbf{Finding \#1:} If you do not account for perspective the ML model will be trained on data points
that were either not correctly classified (up to 1/3 of the vulnerabilities seen in the previous period) or not even available (up to 81.5\% wrt those seen the previous
period).}

\subsection{Models Evaluation}
\label{sec:rq2}
After getting the modified datasets, 
we then applied 5 different ML-based vulnerability-finding tools on both the original dataset and modified datasets and observed the change in their performance. These 5 tools are chosen among different ML tools with different architectures from Table~\ref{tab:sota_ml-based}: Vuldeepecker (RNN/BLSTM), Code2Vec (MLP/AST), ReGVD (GNN, token), CodeBERT and LineVul (Transformers). 
In Table \ref{tab:evaluation_summary_sens_projects}, we list all models' differences between precision (left) and recall (right) results with the \RealTestDataset\xspace and \IdealTestDataset\xspace datasets. For precision, some models work better in the \IdealTestDataset\xspace, and some other work better with \RealTestDataset. While for recall, the models' performances with \IdealTestDataset\xspace datasets are on average higher than with \RealTestDataset\xspace datasets except for CodeBERT in \texttt{linux} dataset, Vuldeepecker in \texttt{NVD Vuldeepecker}, and LineVul/Code2Vec (slightly) in \texttt{openssl dataset}.

We tested whether such a difference between the MLs' performance tested in \RealShort and \IdealShort is significant with the Wilcoxon signed rank test, with effect size estimator for correlated samples by Vargha and Delaney ($A$)~\citep{ruscio2013generalizations}. We believe the $A$ value has here a natural Software Engineering interpretation: the probability that by taking at random a sample with the ``benefit of hindsight'', its ML performance will be higher than the sample performance in reality.
Since we are testing 5 different tools, according to Bonferroni correction, the $p_{value}$ of a test needs to be $< 0.05/5\ or\ 0.01$ to be considered as significant.
The test on recall results in a statistically significant difference for CodeBERT ($T=42.0, p_{value}=0.002, A=0.75$), LineVul ($T=68.0, p_{value}=0.006, A=0.71$), and ReGVD ($T=15.0, p_{value}=0.004, A=0.7$). While for precision, all the tests return insignificant ($0.015 < p_{value} < 0.55$) after the Bonferroni correction.

\begin{table}[t]
    \centering
    \caption{Models evaluation: difference in the precision and recall statistic.}
    \longcaption{\columnwidth}{RF: Random Forest from~\cite{jimenez2019importance}, LineVul, C2V: Code2Vec, CB: CodeBERT, RG: ReGVD, VD: Vuldeepecker. The more positive the gap, the better the performance is. The recall on average is better when evaluated using the \IdealTestDataset\xspace dataset, except for CodeBERT in \texttt{linux} dataset, LineVul and Code2Vec (slightly) in \texttt{openssl} dataset, and Vuldeepecker in \texttt{NVD Vuldeepecker} dataset. This shows that the ML models' \RealTestDataset\xspace performance (\RealShort) can be not as good as we got from testing using the same year data with the training (\IdealShort). }
    \scriptsize
    \resizebox{0.485\textwidth}{!}{
    \begin{tabular}{l|r|rrrrr}
        % \toprule
         \multicolumn{7}{c}{Precision Gap between Perspective and Retrospective Testing} \\ \midrule
        Metric & RF & LV & C2V & CB & RG & VD \\ 
        \hline
        \rowcolor{gray!20}
        \multicolumn{7}{l}{\texttt{Linux}} \\
        \hline

        Mean &  -43.5\% &  +1.3\% &  -0.9\% &  +7.3\% &  +1.0\% &  -0.8\% \\
        St.Dev. &  +31.2\% &  +2.5\% & +4.4\% &  +14.0\% &  +3.3\% &  +2.9\% \\
        \hline
        \rowcolor{gray!20}
        \multicolumn{7}{l}{\texttt{OpenSSL}} \\
        \hline
        Mean &  -8.72\% &  -10.5\% &  -1.2\% &  0.0\% &  +7.2\% &  +1.2\% \\
        St.Dev. &  -25.4\% &  +39.6\% &  +58.4\% &  0.0\% &  +24.4\% &  +13.4\% \\
        \hline
        \rowcolor{gray!20}
        \multicolumn{7}{l}{\texttt{Poppler}} \\
        \hline

        Mean & N/A &  0.0\% &  0.0\% &  0.0\% &  0.0\% &  -2.3\% \\
        St.Dev. & N/A &  0.0\% &  0.0\% &  0.0\% &  0.0\% &  +4.6\% \\
        \hline
        \rowcolor{gray!20}
        \multicolumn{7}{l}{\texttt{NVD Vuldeepecker}} \\
        \hline

        Mean & N/A &  -35.9\% &  -19.8\% &  -10.5\% &  -12.7\% &  +0.7\% \\
        St.Dev. & N/A &  +16.5\% &   +36.2\% &  18.1\% &  +16.4\% &  +9.5\% \\
        \bottomrule
    \end{tabular}}
    \quad
    \resizebox{0.485\textwidth}{!}{
    \begin{tabular}{l|r|rrrrr}
        % \toprule
         \multicolumn{7}{c}{Recall Gap between Perspective and Retrospective Testing} \\ \midrule
        Metric & RF & LV & C2V & CB & RG & VD \\ 
        \hline
        \rowcolor{gray!20}
        \multicolumn{7}{l}{\texttt{Linux}} \\
        \hline
        Mean &  -73.5\% &  -1.4\% &  -3.6\% &  +5.2\% &  -5.0\% &  -14.9\% \\
        St.Dev. &  +18.1\% &  +6.8\% &  +13.3\% &  +14.7\% &  +5.4\% &  +52.2\% \\
        \hline
        \rowcolor{gray!20}
        \multicolumn{7}{l}{\texttt{OpenSSL}} \\
        \hline
        Mean &  -37.0\% &  +0.6\% &  +0.5\% &  -10.1\% &  -4.5\% &  -47.1\% \\
        St.Dev. &  +9.1\% &  +30.9\% &  +9.8\% &  +19.1\% &  +11.3\% &  +49.3\% \\
        \hline
        \rowcolor{gray!20}
        \multicolumn{7}{l}{\texttt{Poppler}} \\
        \hline
        Mean & N/A &  -27.9\% &  0.0\% &  -29.6\% &  0.0\% &  -12.5\% \\
        St.Dev. & N/A &  +40.2\% &  0.0\% &  +15.7\% &  0.0\% &  +25.0\% \\
        \hline
        \rowcolor{gray!20}
        \multicolumn{7}{l}{\texttt{NVD Vuldeepecker}} \\
        \hline
        Mean & N/A &  -47.7\% &  -3.5\% &  -28.5\% &  -33.6\% &  +33.2\% \\
        St.Dev. & N/A &  +29.3\% &  +5.4\% &  -21.0\% &  +34.8\% &  +43.0\% \\
        \bottomrule
        \end{tabular}}
    \label{tab:evaluation_summary_sens_projects}
\end{table}

\begin{figure*}[t]
\vspace{-.2\baselineskip}
    \centering
    \includegraphics[width=0.9\textwidth]{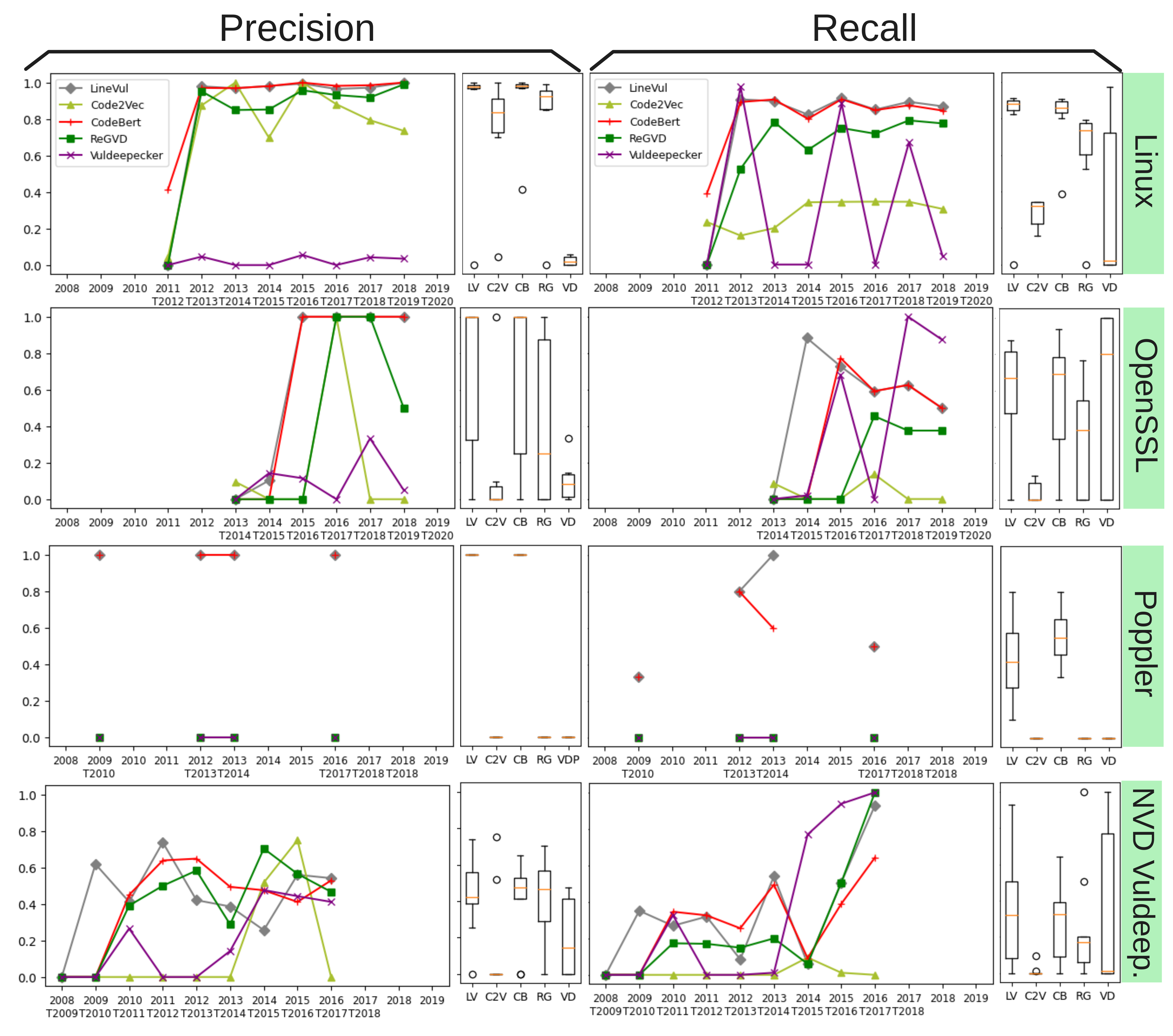}
    \label{fig:prec}
\longcaptionfig{\textwidth}{\scriptsize In \texttt{linux} dataset, the performance of the ML models (except Vuldeepecker) seems to have a trend: it gets better with more retrospective data. However, this trend does not happen in the other 3 datasets.}
\caption{Evolution of Precision and Recall of 5 ML models with more and more retrospectives.}
\label{fig:validation_sens}

\end{figure*}

To see if we can reach better performance by giving more perspective to the model, we portray the precision (left) and recall (right) of the models (tested with \RealTestDataset\xspace dataset) by time in Figure~\ref{fig:validation_sens}. We show line charts to show the trends and boxplots as has been done in~\cite{jimenez2019importance}.
In \texttt{linux} dataset, most models, except Vuldeepecker, have on average high precision and recall. However, there are no increasing trends as one could hypothesize with more and more retrospective data. Vuldeepecker's recall even goes up and down, showing no trend. Interestingly, in the \texttt{OpenSSL} dataset, the figure changes. LineVul, ReGVD, and CodeBERT still have high precision, but their recall drops with more and more retrospectives, while Code2Vec and Vuldeepecker show no trend by going up and down.

While LineVul seems to work well on \texttt{linux} dataset, it performs worse in the \texttt{NVD Vuldeepecker} dataset (lower right) with lower precision and recall (average 34.01\% st.dev. 27.46\%). Not only LineVul, ReGVD, and CodeBERT also have on average lower precision and recall in this dataset 
These three models have an average low recall until 2014 but then their recall increases significantly in 2015 and 2016. Most likely at that time, they are trained with enough retrospective information to identify vulnerability in the next year. Vuldeepecker has a dual behavior from the paper to the field setting. It also has an increasing recall trend from 2013 to 2016. Unlike Vuldeepecker, Code2Vec keeps failing to identify vulnerability as it did in the \IdealShort case.

We tested if the models exhibited a positive (increasing) trend of performance across the years (\RealDataset) with the Mann-Kendall trend test. We ran the Mann-Kendall test for each model in the 4 datasets. According to Bonferroni correction, the $p_{value}$ of a test needs to be $< 0.05/4\ or\ 0.0125$ to be considered as significant. As shown in Table~\ref{tab:mann-kendall}, none of the tests of the precision returns significant, which means we cannot conclude any increasing (or decreasing) trend in the precision of the models when tested in unknown future data, even with more and more retrospective. For the recall, only Vuldeepecker on its own dataset (NVD Vuldeepecker) returns a significant increasing trend. 

\begin{table}[]
    \centering
    \scriptsize
    \caption{Mann-Kendall Trend Test on the ML Performance Tested on Perspective Data}
    \longcaption{0.92\textwidth}{We did the Mann-Kendall test for each model in the 4 datasets, therefore, $p_{value}$ needs to be less than $0.0125$ for the test to be significant. There is only one significant increasing trend: Vuldeepecker in their own dataset (NVD Vuldeepecker).}
    \begin{tabular}{ll|lrr}
    \multicolumn{5}{c}{Precision}\\
    \hline
        Model & Dataset & Trend & MK & $p$\\
        \hline
        & Linux & No trend & 12 & 0.17 \\
         & OpenSSL & No trend & 9 & 0.07 \\
         & Poppler & No trend & 0 & 1.00 \\
         \multirow{-4}{*}{LineVul}& NVD Vuldeep. & No trend & 2 & 0.92 \\\hline
         & Linux & No trend & 1 & 1.00 \\
         & OpenSSL & No trend & -3 & 0.65 \\
         & Poppler & No trend & 0 & 1.00 \\
        \multirow{-4}{*}{Code2Vec} & NVD Vuldeep. & No trend & 11 & 0.15 \\\hline
         & Linux & No trend & 21 & 0.01 \\
         & OpenSSL & No trend & 8 & 0.11 \\
         & Poppler & No trend & 0 & 1.00 \\
          \multirow{-4}{*}{CodeBERT} & NVD Vuldeep. & No trend & 11 & 0.29 \\\hline
         & Linux & No trend & 14 & 0.11 \\
         & OpenSSL & No trend & 7 & 0.22 \\
         & Poppler & No trend & 0 & 1.00 \\
         \multirow{-4}{*}{ReGVD} & NVD Vuldeep. & No trend & 17 & 0.09 \\\hline
         & Linux & No trend & 4 & 0.69 \\
         & OpenSSL & No trend & 2 & 0.85 \\
         & Poppler & No trend & 0 & 1.00 \\
         \multirow{-4}{*}{Vuldeepecker}& NVD Vuldeep. & No trend & 18 & 0.06 \\
        \hline
    \end{tabular}
    \quad
    \begin{tabular}{ll|lrr}
    \multicolumn{5}{c}{Recall}\\
    \hline
        Dataset & Model & Trend & MK & $p$\\
        \hline
        & Linux & No trend & 2 & 0.90 \\
         & OpenSSL & No trend & -3 & 0.71 \\
         & Poppler & No trend & 2 & 0.73 \\
         \multirow{-4}{*}{LineVul} & NVD Vuldeep. & No trend & 14 & 0.18 \\\hline
        & Linux & No trend & 14 & 0.11 \\
         & OpenSSL & No trend & -3 & 0.65 \\
         & Poppler & No trend & 0 & 1.0 \\
         \multirow{-4}{*}{Code2Vec} & NVD Vuldeep. & No trend & 9 & 0.25 \\ \hline
         & Linux & No trend & 2 & 0.90 \\
         & OpenSSL & No trend & 4 & 0.57 \\
         & Poppler & No trend & 0 & 1.00 \\
         \multirow{-4}{*}{CodeBERT} & NVD Vuldeep. & No trend & 19 & 0.06 \\\hline
         & Linux & No trend & 16 & 0.06 \\
         & OpenSSL & No trend & 7 & 0.22 \\
         & Poppler & No trend & 0 & 1.00 \\
         \multirow{-4}{*}{ReGVD} & NVD Vuldeep. & No trend & 21 & 0.04 \\\hline
         & Linux & No trend & 2 & 0.89 \\
         & OpenSSL & No trend & 8 & 0.18 \\
         & Poppler & No trend & 0 & 1.00 \\
         \multirow{-4}{*}{Vuldeepecker}  & NVD Vuldeep. & \cellcolor{green!20}Increasing & 24 & 0.01 \\
        \hline
    \end{tabular}
    \label{tab:mann-kendall}
\end{table}

\highlight{black}{\textbf{Finding \#2:} Observation over the years shows that there is no consistent ML performance of the prediction when tested in the \RealShort case.
}

\section{Additional Extensions}
\label{sec:ext}
We also ran additional experiments with \textit{seen but believed negatives}. These data points represent the codes available at a certain time of observation $t$ and \textit{considered as} not vulnerable but are actually found vulnerable in the future. These data points are negatives at time $t$, but then become positive at $t+1$. Assuming that the data (the code) exists beforehand, we add these positive points as negatives to our retrospective test set. 
The result of this additional experiment shows that the recall does not change for most models and datasets, but the precision dropped. This happens because the models are still classifying most of the \textit{believed negative} data as vulnerable. This high False Positive at a certain time $t$ would be considered as bad, but they are actually classifying the code correctly (as vulnerable/ positive), just the label was still different at time $t$.

\section{Threat to Validity and Future Works}
\label{sec:tov}
\subsection{Validation Case}
For our preliminary validation, we chose vulnerability detection as a case study.
We chose this case study as vulnerability detection has been a popular field, but the ML model's performance still needs to be improved to find vulnerabilities in a real-world setting~\cite{chakraborty2021deep}. However, we believe that our methodology can be applied to any ML model evaluation with a time-based dataset, and we plan to do more evaluations with other cases in the future.

\subsection{Validation Dataset}
We only applied our methodology to 4 datasets (3 extracted from BigVul~\cite{fan2020ac} and 1 from Vuldeepecker~\cite{li_vuldeepecker_2018}) in our validation. 

\paragraph{Quality.}
There is a possibility that the chosen datasets may have some wrong labels~\cite{croft2023data}. Obviously, the quality does indeed affect the ML algorithm's performance. Our focus is to show that no matter the dataset or model, if one still tests on the entire dataset (retrospectively), one will get a better result than one would obtain in the field. The wrong labels would therefore not impact the phenomenon we want to measure: they would be wrong when considered during the retrospective analysis, the properly timed training phase, and the properly timed testing phase. Their bias will therefore be uniform across all cases considered in this paper.

\paragraph{Completeness.}
The chosen datasets have a limitation in that they do not have information on the code availability date. Therefore, we use labeling dates as a proxy for availability dates during implementation. We acknowledge the limitation of this proxy as a possible threat to validity. However, as the assumption that the availability dates are less than or equal to the labeling dates holds, we argue that this proxy is sufficient to show the difference between the retrospective and perspective views. 

\paragraph{Future works.}
In the future, we plan to validate it using other datasets, such as the ProjectKB~\cite{ponta2019msr} dataset (which has not been evaluated for ML vulnerability detection but contains mostly Java code and vulnerabilities) to see how the method works for another language. We believe that our methodology can be applied to any time-based dataset, language-agnostic, and we encourage the research community to develop and use more time-based datasets to validate the ML models in the perspective vs. retrospective setting.

\subsection{Validation Models}
We only validate our methodology by running 5 ML models: Vuldeepecker~\cite{li_vuldeepecker_2018}, Code2Vec \cite{alon2019code2vec}, ReGVD~\cite{nguyen2022regvd}, CodeBERT~\cite{feng2020codebert}, and LineVul~\cite{fu2022linevul}. While any comparison is of course limited, this selection covered a sufficient variety of different ML techniques used (RNN/BLSTM, MLP/AST, GNN, and Transformers).

\paragraph{LLMs.}
We did not include LLMs in this study because ours is a comparison between retrospective and prospective studies. As of today, we can only run a prospective study with LLMs, because available pre-trained LLMs already know the past, and we cannot ``remove the past'' from them. Unlike other ML-based models, which can be trained from scratch on a given dataset, we cannot retrain LLMs from scratch because we do not have a snapshot of the entire internet in 2022.
To be able to use an LLM, we must use an LLM pre-trained in 2021 and test on vulnerability data of 2022, then take an LLM pre-trained in 2022 and test on 2023, and so on. As of today, we do not have enough temporal data to make robust conclusions, but it would be an interesting future study to be done 2 years from now with at least 5 years of observation.

\paragraph{Only 5 models: future replication.}
Even if we cannot include all available models, our methodology is general and applicable to other ML models. Moreover, the trend in the performance of each evaluated model consistently shows a difference between performance with full retrospective information and the performance at a point in time where retrospective information is unavailable. Testing with more models will become a replication study in future work.

\subsection{Validation Metrics}
We decided to use precision and recall as metrics to show the comparison among the results. 
One can choose to use another metric, but all metrics are constructed from basic metrics: TP, FP, TN, and FN, which are dependent on each other, i.e., when the TP increases, the FN will decrease, etc.
This resulted in metrics that are interrelated with each other by a linear equation.
Of course, it is still possible that in particular cases, the precision and recall \emph{always} cancel each other, which removes the fluctuation from F1. However, these cases are extremely unlikely in complex experiments with ML tools and real data.

\subsection{Reasons of Fluctuating Trends}
In this paper, we only showed and compared the trends between \textit{Retrospective Testing (R-R)} and \textit{Perspective Testing (R-P)}. We did not investigate further the reasons why there is no trend and why the results fluctuate over the years. 
Yet, our goal is to show that \textbf{this is a problem} one will encounter in reality. Models overfitting and intrinsic or seasonal trends (a known phenomenon since~\cite{joh2009seasonal}) can all be subject to further investigation. As far as we know, research on ML result stability in software engineering is limited and remains a future research direction.  

\section{Implications of the Findings} \label{sec:implications}

\begin{table}[t]
\footnotesize
  \caption{Summary of Findings and Implications.}
  \vspace{-.5\baselineskip}
  \label{tab:implications}
  \resizebox{\columnwidth}{!}{
  \begin{tabular}{l|p{0.37\textwidth}|p{0.53\textwidth}}
    \hline
    RQ & Main Finding & Implication for Research\\
    \hline
    RQ1 & If you do not account for perspective the ML model will be trained on data points that were either not correctly classified (up to 1/3 of the vulnerabilities seen in the previous period) or not even available (up to 81.5\% wrt those seen the previous period) & Today's cat is tomorrow's dog: label of codes can change over time as new vulnerabilities are being found, both in training and testing set. The ML models’ performance on retrospective (\Ideal) might not fully capture performance on perspective (\Real). \\
    RQ2 &  Observation over the years (Figure~\ref{fig:validation_sens}) and Mann-Kendall test results show no consistent ML performance of the prediction when tested in the \RealShort case. & Presenting ML's performances in time as \emph{trends} will provide a more realistic understanding, as it also shows how much retrospective information affects ML's performance.\\
  \hline
    \end{tabular}}
\end{table}
Table~\ref{tab:implications} summarizes the implications and the main findings from each research question.
Our first result (\S\ref{sec:rq1}) shows that ML models' performance could (most probably) be different when used in the \textit{Perspective Testing (R-P)} compared to the testing result which in most papers is in the \textit{Retrospective Testing (R-R)}. This result on ML for source code is aligned with the result by~\cite{jimenez2019importance} on code metrics. Research evaluating ML performance should consider adopting our methodology to consider perspective information in their evaluation and present results that reflect more than one scenario at a given point in time.
Our result in RQ2 (\S\ref{sec:rq2}) shows that the trends are informative in visualizing the impact of retrospective information on ML performance. 

A key issue to debate is our choice of \emph{not} using the complete information for the testing dataset. The motivation behind this choice is that vulnerabilities are discovered sometimes \emph{after years} from the release. This has been a consistent finding since the milk or wine study \cite{ozment2006milk}, which introduced the notion of foundational vulnerability (present since the very beginning of a project). Retrospective discovery is common across systems \cite{nguyen2016automatic} and ecosystems \cite{hu2024empirical}. Even looking at the famous log4j vulnerability, CVE-2021-4104 refers to a version that reached EoL in 2015, an after-life vulnerability \cite{massacci2011after}. Consider a developer running a model on log4j in 2015. Even if by mistake the model flagged the vulnerable fragment (there was no evidence of similar vulnerabilities at the time), it would have been considered a false positive. The model would have been considered a failure for five long years.

\section{Conclusions}
\label{sec:conclusion}

In this work, we propose a methodology to produce a timeline of partial, time-actual information datasets from a complete information (retrospective) dataset. For each time point in the timeline, we generate a training set and 2 testing sets: one simulating performance on the \Ideal (testing with the information available until the chosen time point) and one simulating the performance on the \Real (testing until the next time point). The former corresponds to the case of a researcher arguing on paper for the model to be deployed based on the available information at the time, the latter corresponds to performance experienced (or perceived) on the field for the newly developed software, which will be classified by the ML model until the next time point.

The key, real-world issue we try to capture with this work is the fact that labels change over time. This is probably the most critical difference with datasets used for image classification, where most ML algorithms have been developed. An image of a `true' cat will never become an image of a `true' dog. The application of ML methods to software engineering and security vulnerability detection, in particular, must take this difference into account when evaluating models. 

We validated the methodology by using time slots that are one year apart from each other. The resulting test datasets show that the number of vulnerabilities an ML model has to identify can be really different from the ones they are tested with. This finding supports our next finding: when tested using \RealTestDataset\xspace datasets, the ML models performed worse when compared to the results from using the \IdealTestDataset\xspace dataset. This result is partly aligned with the results found in~\cite{jimenez2019importance} on code metrics, which broke the full information available to training by releases, but still uses the full information to test the success of models trained on partial information. 

From these findings, we want to raise awareness of the impact of retrospective information when evaluating ML models with any time-based dataset and the possibility of evaluating ML models using perspective datasets instead. We also found that ML models have no significant trend as more and more retrospective information is added, as one might expect, as shown by the results of the Mann-Kendall tests and the visual representation of the results.
We believe that our methodology also applies to other use cases, e.g., bug/ defect detection and commit classification, therefore, we plan to do more validation in the future with different datasets and ML models and use cases.

\section*{Acknowledgments}
This work was partly funded by the EU under the H2020 Program AssureMOSS (Grant n. 952647) and the Horizon Europe Program Sec4AI4Sec (Grant n. 101120393), by the Italian Ministry of University and Research (MUR) under the P.N.R.R. – NextGenerationEU grant n.\ PE00000014 (SERICS subproject COVERT), and by the Dutch Research Council (NWO) under the grant NWA.1215.18.006 (Theseus) and grant KIC1.VE01.20.004 (HEWSTI).

\section*{Credit Statements.}
\noindent \emph{Conceptualization:} RP, FM; %Ideas; formulation or evolution of overarching research goals and aims
	\emph{Methodology:} RP, FM; 	%Development or design of methodology; creation of models
 
\noindent \emph{Software:} RP, YF; 	%Programming, software development; designing computer programs; implementation of the computer code and supporting algorithms; testing of existing code components
	\emph{Validation:} RP, YF;	%Verification, whether as a part of the activity or separate, of the overall replication/ reproducibility of results/experiments and other research outputs
 
\noindent \emph{Formal analysis:} RP, YF, FM;	%Application of statistical, mathematical, computational, or other formal techniques to analyze or synthesize study data
\emph{Investigation:} RP, YF;	%Conducting a research and investigation process, specifically performing the experiments, or data/evidence collection
    
\noindent \emph{Data Curation:} RP, YF; 	%Management activities to annotate (produce metadata), scrub data and maintain research data (including software code, where it is necessary for interpreting the data itself) for initial use and later reuse
\emph{Writing - Original Draft:} RP, YF;	%Preparation, creation and/or presentation of the published work, specifically writing the initial draft (including substantive translation)
    
\noindent \emph{Writing - Review \& Editing:} RP, YF, FM;	%Preparation, creation and/or presentation of the published work by those from the original research group, specifically critical review, commentary or revision – including pre-or postpublication stages
    
\noindent \emph{Visualization:} RP;	%Preparation, creation and/or presentation of the published work, specifically visualization/ data presentation
\emph{Supervision:} FM;	%Oversight and leadership responsibility for the research activity planning and execution, including mentorship external to the core team
    
\noindent \emph{Project administration:} FM;	%Management and coordination responsibility for the research activity planning and execution
\emph{Funding acquisition:} FM;	%Acquisition of the financial support for the project leading to this publication

\section{Data Availability Statement}
We made available the replication packages of this work in Zenodo~\cite{zenodo-us-1,zenodo-us-2,zenodo-us-3,zenodo-us-4}. There are four repositories, each one for each dataset we used to validate our methodologies: one~\cite{zenodo-us-1} from NVD Vuldeepecker~\cite{li_vuldeepecker_2018} and three from BigVul~\cite{fan2020ac}: \texttt{linux}~\cite{zenodo-us-2}, \texttt{openSSL}~\cite{zenodo-us-3}, and \texttt{poppler}~\cite{zenodo-us-4}.
Each replication package includes: (1) \emph{Code} that (a) implements our methodology to generate the datasets, (b) runs the models in our validation, and (c) generates charts from the ML evaluation results, (2) \emph{Datasets} contains (a) the original datasets from Vuldeepecker~\cite{li_vuldeepecker_2018} and BigVul~\cite{fan2020ac} and (b) the datasets we created using our methodology, (3) \emph{Pretrained-models} that we generated during our evaluation, and (4) \emph{Results} of our evaluation.

\bibliographystyle{ACM-Reference-Format}
\bibliography{references}


\begin{thebibliography}{69}

%%% the Web address.

\ifx \showCODEN    \undefined \def \showCODEN     #1{\unskip}     \fi
\ifx \showISBNx    \undefined \def \showISBNx     #1{\unskip}     \fi
\ifx \showISBNxiii \undefined \def \showISBNxiii  #1{\unskip}     \fi
\ifx \showISSN     \undefined \def \showISSN      #1{\unskip}     \fi
\ifx \showLCCN     \undefined \def \showLCCN      #1{\unskip}     \fi
\ifx \shownote     \undefined \def \shownote      #1{#1}          \fi
\ifx \showarticletitle \undefined \def \showarticletitle #1{#1}   \fi
\ifx \showURL      \undefined \def \showURL       {\relax}        \fi
% invisible to TeX
\providecommand\bibfield[2]{#2}
\providecommand\bibinfo[2]{#2}
\providecommand\natexlab[1]{#1}
\providecommand\showeprint[2][]{arXiv:#2}

\bibitem[Ahmad et~al\mbox{.}(2021)]%
        {ahmad2021unified}
\bibfield{author}{\bibinfo{person}{Wasi Ahmad}, \bibinfo{person}{Saikat Chakraborty}, \bibinfo{person}{Baishakhi Ray}, {and} \bibinfo{person}{Kai-Wei Chang}.} \bibinfo{year}{2021}\natexlab{}.
\newblock \showarticletitle{Unified Pre-training for Program Understanding and Generation}. In \bibinfo{booktitle}{\emph{Proceedings of the 2021 Conference of the North American Chapter of the Association for Computational Linguistics: Human Language Technologies}}, \bibfield{editor}{\bibinfo{person}{Kristina Toutanova}, \bibinfo{person}{Anna Rumshisky}, \bibinfo{person}{Luke Zettlemoyer}, \bibinfo{person}{Dilek Hakkani-Tur}, \bibinfo{person}{Iz~Beltagy}, \bibinfo{person}{Steven Bethard}, \bibinfo{person}{Ryan Cotterell}, \bibinfo{person}{Tanmoy Chakraborty}, {and} \bibinfo{person}{Yichao Zhou}} (Eds.). \bibinfo{publisher}{Association for Computational Linguistics}, \bibinfo{address}{Online}, \bibinfo{pages}{2655--2668}.
\newblock
\href{https://doi.org/10.18653/v1/2021.naacl-main.211}{doi:\nolinkurl{10.18653/v1/2021.naacl-main.211}}


\bibitem[Alon et~al\mbox{.}(2019)]%
        {alon2019code2vec}
\bibfield{author}{\bibinfo{person}{Uri Alon}, \bibinfo{person}{Meital Zilberstein}, \bibinfo{person}{Omer Levy}, {and} \bibinfo{person}{Eran Yahav}.} \bibinfo{year}{2019}\natexlab{}.
\newblock \showarticletitle{code2vec: Learning distributed representations of code}.
\newblock \bibinfo{journal}{\emph{Proceedings of the ACM on Programming Languages}} \bibinfo{volume}{3}, \bibinfo{number}{POPL} (\bibinfo{year}{2019}), \bibinfo{pages}{1--29}.
\newblock
\href{https://doi.org/10.1145/3290353}{doi:\nolinkurl{10.1145/3290353}}


\bibitem[Arp et~al\mbox{.}(2022)]%
        {arp2022and}
\bibfield{author}{\bibinfo{person}{Daniel Arp}, \bibinfo{person}{Erwin Quiring}, \bibinfo{person}{Feargus Pendlebury}, \bibinfo{person}{Alexander Warnecke}, \bibinfo{person}{Fabio Pierazzi}, \bibinfo{person}{Christian Wressnegger}, \bibinfo{person}{Lorenzo Cavallaro}, {and} \bibinfo{person}{Konrad Rieck}.} \bibinfo{year}{2022}\natexlab{}.
\newblock \showarticletitle{Dos and don'ts of machine learning in computer security}. In \bibinfo{booktitle}{\emph{31st USENIX Security Symposium (USENIX Security 22)}}. \bibinfo{pages}{3971--3988}.
\newblock
\href{https://doi.org/10.48550/arXiv.2010.09470}{doi:\nolinkurl{10.48550/arXiv.2010.09470}}


\bibitem[Bhandari et~al\mbox{.}(2021)]%
        {bhandari2021cvefixes}
\bibfield{author}{\bibinfo{person}{Guru Bhandari}, \bibinfo{person}{Amara Naseer}, {and} \bibinfo{person}{Leon Moonen}.} \bibinfo{year}{2021}\natexlab{}.
\newblock \showarticletitle{CVEfixes: automated collection of vulnerabilities and their fixes from open-source software}. In \bibinfo{booktitle}{\emph{Proceedings of the 17th International Conference on Predictive Models and Data Analytics in Software Engineering}}. \bibinfo{pages}{30--39}.
\newblock
\href{https://doi.org/10.1145/3475960.3475985}{doi:\nolinkurl{10.1145/3475960.3475985}}


\bibitem[Chakraborty et~al\mbox{.}(2021)]%
        {chakraborty2021deep}
\bibfield{author}{\bibinfo{person}{Saikat Chakraborty}, \bibinfo{person}{Rahul Krishna}, \bibinfo{person}{Yangruibo Ding}, {and} \bibinfo{person}{Baishakhi Ray}.} \bibinfo{year}{2021}\natexlab{}.
\newblock \showarticletitle{Deep learning based vulnerability detection: Are we there yet}.
\newblock \bibinfo{journal}{\emph{IEEE Transactions on Software Engineering}} (\bibinfo{year}{2021}).
\newblock
\href{https://doi.org/10.1109/TSE.2021.3087402}{doi:\nolinkurl{10.1109/TSE.2021.3087402}}


\bibitem[Chen et~al\mbox{.}(2019)]%
        {chen2019using}
\bibfield{author}{\bibinfo{person}{Haipeng Chen}, \bibinfo{person}{Rui Liu}, \bibinfo{person}{Noseong Park}, {and} \bibinfo{person}{VS Subrahmanian}.} \bibinfo{year}{2019}\natexlab{}.
\newblock \showarticletitle{Using twitter to predict when vulnerabilities will be exploited}. In \bibinfo{booktitle}{\emph{Proceedings of the 25th ACM SIGKDD international conference on knowledge discovery \& data Mining}}. \bibinfo{pages}{3143--3152}.
\newblock
\href{https://doi.org/10.1145/3292500.3330742}{doi:\nolinkurl{10.1145/3292500.3330742}}


\bibitem[Chen et~al\mbox{.}(2023)]%
        {chen2023diversevul}
\bibfield{author}{\bibinfo{person}{Yizheng Chen}, \bibinfo{person}{Zhoujie Ding}, \bibinfo{person}{Lamya Alowain}, \bibinfo{person}{Xinyun Chen}, {and} \bibinfo{person}{David Wagner}.} \bibinfo{year}{2023}\natexlab{}.
\newblock \showarticletitle{DiverseVul: A New Vulnerable Source Code Dataset for Deep Learning Based Vulnerability Detection}. In \bibinfo{booktitle}{\emph{Proceedings of the 26th International Symposium on Research in Attacks, Intrusions and Defenses}} (Hong Kong, China) \emph{(\bibinfo{series}{RAID '23})}. \bibinfo{publisher}{Association for Computing Machinery}, \bibinfo{address}{New York, NY, USA}, \bibinfo{pages}{654–668}.
\newblock
\showISBNx{9798400707650}
\href{https://doi.org/10.1145/3607199.3607242}{doi:\nolinkurl{10.1145/3607199.3607242}}


\bibitem[Croft et~al\mbox{.}(2023)]%
        {croft2023data}
\bibfield{author}{\bibinfo{person}{Roland Croft}, \bibinfo{person}{M~Ali Babar}, {and} \bibinfo{person}{M~Mehdi Kholoosi}.} \bibinfo{year}{2023}\natexlab{}.
\newblock \showarticletitle{Data quality for software vulnerability datasets}. In \bibinfo{booktitle}{\emph{2023 IEEE/ACM 45th International Conference on Software Engineering (ICSE)}}. IEEE, \bibinfo{pages}{121--133}.
\newblock
\href{https://doi.org/10.1109/ICSE48619.2023.00022}{doi:\nolinkurl{10.1109/ICSE48619.2023.00022}}


\bibitem[Dallmeier and Zimmermann(2007)]%
        {dallmeier2007extraction}
\bibfield{author}{\bibinfo{person}{Valentin Dallmeier} {and} \bibinfo{person}{Thomas Zimmermann}.} \bibinfo{year}{2007}\natexlab{}.
\newblock \showarticletitle{Extraction of Bug Localization Benchmarks from History}. In \bibinfo{booktitle}{\emph{Proceedings of the 22nd IEEE/ACM International Conference on Automated Software Engineering}} (Atlanta, Georgia, USA) \emph{(\bibinfo{series}{ASE '07})}. \bibinfo{publisher}{Association for Computing Machinery}, \bibinfo{address}{New York, NY, USA}, \bibinfo{pages}{433–436}.
\newblock
\showISBNx{9781595938824}
\href{https://doi.org/10.1145/1321631.1321702}{doi:\nolinkurl{10.1145/1321631.1321702}}


\bibitem[De~Sousa and Hasselbring(2021)]%
        {desousa2021javabert}
\bibfield{author}{\bibinfo{person}{Nelson~Tavares De~Sousa} {and} \bibinfo{person}{Wilhelm Hasselbring}.} \bibinfo{year}{2021}\natexlab{}.
\newblock \showarticletitle{JavaBERT: Training a Transformer-Based Model for the Java Programming Language}. In \bibinfo{booktitle}{\emph{2021 36th IEEE/ACM International Conference on Automated Software Engineering Workshops (ASEW)}}. \bibinfo{pages}{90--95}.
\newblock
\href{https://doi.org/10.1109/ASEW52652.2021.00028}{doi:\nolinkurl{10.1109/ASEW52652.2021.00028}}


\bibitem[dos Santos and Figueiredo(2020)]%
        {dos2020commit}
\bibfield{author}{\bibinfo{person}{Geanderson~E dos Santos} {and} \bibinfo{person}{Eduardo Figueiredo}.} \bibinfo{year}{2020}\natexlab{}.
\newblock \showarticletitle{Commit Classification using Natural Language Processing: Experiments over Labeled Datasets.}. In \bibinfo{booktitle}{\emph{CIbSE}}. \bibinfo{pages}{110--123}.
\newblock


\bibitem[Fan et~al\mbox{.}(2020)]%
        {fan2020ac}
\bibfield{author}{\bibinfo{person}{Jiahao Fan}, \bibinfo{person}{Yi Li}, \bibinfo{person}{Shaohua Wang}, {and} \bibinfo{person}{Tien~N Nguyen}.} \bibinfo{year}{2020}\natexlab{}.
\newblock \showarticletitle{A C/C++ code vulnerability dataset with code changes and CVE summaries}. In \bibinfo{booktitle}{\emph{Proceedings of the 17th International Conference on Mining Software Repositories}}. \bibinfo{pages}{508--512}.
\newblock
\href{https://doi.org/10.1145/3379597.3387501}{doi:\nolinkurl{10.1145/3379597.3387501}}


\bibitem[Feng et~al\mbox{.}(2020)]%
        {feng2020codebert}
\bibfield{author}{\bibinfo{person}{Zhangyin Feng}, \bibinfo{person}{Daya Guo}, \bibinfo{person}{Duyu Tang}, \bibinfo{person}{Nan Duan}, \bibinfo{person}{Xiaocheng Feng}, \bibinfo{person}{Ming Gong}, \bibinfo{person}{Linjun Shou}, \bibinfo{person}{Bing Qin}, \bibinfo{person}{Ting Liu}, \bibinfo{person}{Daxin Jiang}, {and} \bibinfo{person}{Ming Zhou}.} \bibinfo{year}{2020}\natexlab{}.
\newblock \showarticletitle{{C}ode{BERT}: A Pre-Trained Model for Programming and Natural Languages}. In \bibinfo{booktitle}{\emph{Findings of the Association for Computational Linguistics: EMNLP 2020}}, \bibfield{editor}{\bibinfo{person}{Trevor Cohn}, \bibinfo{person}{Yulan He}, {and} \bibinfo{person}{Yang Liu}} (Eds.). \bibinfo{publisher}{Association for Computational Linguistics}, \bibinfo{address}{Online}, \bibinfo{pages}{1536--1547}.
\newblock
\href{https://doi.org/10.18653/v1/2020.findings-emnlp.139}{doi:\nolinkurl{10.18653/v1/2020.findings-emnlp.139}}


\bibitem[Fu and Tantithamthavorn(2022)]%
        {fu2022linevul}
\bibfield{author}{\bibinfo{person}{Michael Fu} {and} \bibinfo{person}{Chakkrit Tantithamthavorn}.} \bibinfo{year}{2022}\natexlab{}.
\newblock \showarticletitle{Linevul: A transformer-based line-level vulnerability prediction}. In \bibinfo{booktitle}{\emph{Proceedings of the 19th International Conference on Mining Software Repositories}}. \bibinfo{pages}{608--620}.
\newblock
\href{https://doi.org/10.1145/3524842.3528452}{doi:\nolinkurl{10.1145/3524842.3528452}}


\bibitem[Ghadhab et~al\mbox{.}(2021)]%
        {ghadhab2021augmenting}
\bibfield{author}{\bibinfo{person}{Lobna Ghadhab}, \bibinfo{person}{Ilyes Jenhani}, \bibinfo{person}{Mohamed~Wiem Mkaouer}, {and} \bibinfo{person}{Montassar~Ben Messaoud}.} \bibinfo{year}{2021}\natexlab{}.
\newblock \showarticletitle{Augmenting commit classification by using fine-grained source code changes and a pre-trained deep neural language model}.
\newblock \bibinfo{journal}{\emph{Information and Software Technology}}  \bibinfo{volume}{135} (\bibinfo{year}{2021}), \bibinfo{pages}{106566}.
\newblock
\href{https://doi.org/10.1016/j.infsof.2021.106566}{doi:\nolinkurl{10.1016/j.infsof.2021.106566}}


\bibitem[Hanif and Maffeis(2022)]%
        {hanif2022vulberta}
\bibfield{author}{\bibinfo{person}{Hazim Hanif} {and} \bibinfo{person}{Sergio Maffeis}.} \bibinfo{year}{2022}\natexlab{}.
\newblock \showarticletitle{Vulberta: Simplified source code pre-training for vulnerability detection}. In \bibinfo{booktitle}{\emph{2022 International joint conference on neural networks (IJCNN)}}. IEEE, \bibinfo{pages}{1--8}.
\newblock
\href{https://doi.org/10.1109/IJCNN55064.2022.9892280}{doi:\nolinkurl{10.1109/IJCNN55064.2022.9892280}}


\bibitem[Hu et~al\mbox{.}(2024)]%
        {hu2024empirical}
\bibfield{author}{\bibinfo{person}{Jinchang Hu}, \bibinfo{person}{Lyuye Zhang}, \bibinfo{person}{Chengwei Liu}, \bibinfo{person}{Sen Yang}, \bibinfo{person}{Song Huang}, {and} \bibinfo{person}{Yang Liu}.} \bibinfo{year}{2024}\natexlab{}.
\newblock \showarticletitle{Empirical Analysis of Vulnerabilities Life Cycle in Golang Ecosystem}. In \bibinfo{booktitle}{\emph{Proceedings of the IEEE/ACM 46th International Conference on Software Engineering}}. \bibinfo{pages}{1--13}.
\newblock
\href{https://doi.org/10.1145/3597503.3639230}{doi:\nolinkurl{10.1145/3597503.3639230}}


\bibitem[Jimenez et~al\mbox{.}(2018)]%
        {jimenez2018engineering}
\bibfield{author}{\bibinfo{person}{Matthieu Jimenez}, \bibinfo{person}{Yves Le~Traon}, {and} \bibinfo{person}{Mike Papadakis}.} \bibinfo{year}{2018}\natexlab{}.
\newblock \showarticletitle{[Engineering Paper] Enabling the Continuous Analysis of Security Vulnerabilities with VulData7}. In \bibinfo{booktitle}{\emph{2018 IEEE 18th International Working Conference on Source Code Analysis and Manipulation (SCAM)}}. IEEE, \bibinfo{pages}{56--61}.
\newblock
\href{https://doi.org/10.1109/SCAM.2018.00014}{doi:\nolinkurl{10.1109/SCAM.2018.00014}}


\bibitem[Jimenez et~al\mbox{.}(2016)]%
        {Jimenez_2016}
\bibfield{author}{\bibinfo{person}{Matthieu Jimenez}, \bibinfo{person}{Mike Papadakis}, {and} \bibinfo{person}{Yves~Le Traon}.} \bibinfo{year}{2016}\natexlab{}.
\newblock \showarticletitle{Vulnerability Prediction Models: A Case Study on the Linux Kernel}. In \bibinfo{booktitle}{\emph{2016 IEEE 16th International Working Conference on Source Code Analysis and Manipulation (SCAM)}}. \bibinfo{pages}{1--10}.
\newblock
\href{https://doi.org/10.1109/SCAM.2016.15}{doi:\nolinkurl{10.1109/SCAM.2016.15}}


\bibitem[Jimenez et~al\mbox{.}(2019)]%
        {jimenez2019importance}
\bibfield{author}{\bibinfo{person}{Matthieu Jimenez}, \bibinfo{person}{Renaud Rwemalika}, \bibinfo{person}{Mike Papadakis}, \bibinfo{person}{Federica Sarro}, \bibinfo{person}{Yves Le~Traon}, {and} \bibinfo{person}{Mark Harman}.} \bibinfo{year}{2019}\natexlab{}.
\newblock \showarticletitle{The importance of accounting for real-world labelling when predicting software vulnerabilities}. In \bibinfo{booktitle}{\emph{Proceedings of the 2019 27th ACM Joint Meeting on European Software Engineering Conference and Symposium on the Foundations of Software Engineering}}. \bibinfo{pages}{695--705}.
\newblock
\href{https://doi.org/10.1145/3338906.3338941}{doi:\nolinkurl{10.1145/3338906.3338941}}


\bibitem[Joh and Malaiya(2009)]%
        {joh2009seasonal}
\bibfield{author}{\bibinfo{person}{HyunChul Joh} {and} \bibinfo{person}{Yashwant~K Malaiya}.} \bibinfo{year}{2009}\natexlab{}.
\newblock \showarticletitle{Seasonal variation in the vulnerability discovery process}. In \bibinfo{booktitle}{\emph{2009 International Conference on Software Testing Verification and Validation}}. IEEE, \bibinfo{pages}{191--200}.
\newblock
\href{https://doi.org/10.1109/ICST.2009.9}{doi:\nolinkurl{10.1109/ICST.2009.9}}


\bibitem[Johnb110(2022)]%
        {vdpython}
\bibfield{author}{\bibinfo{person}{Johnb110}.} \bibinfo{year}{2022}\natexlab{}.
\newblock \bibinfo{title}{VDPython}.
\newblock \bibinfo{howpublished}{\textsc{url:}~\url{https://github.com/johnb110/VDPython}}.
\newblock


\bibitem[Just et~al\mbox{.}(2014)]%
        {defects4j}
\bibfield{author}{\bibinfo{person}{Ren{\'e} Just}, \bibinfo{person}{Darioush Jalali}, {and} \bibinfo{person}{Michael~D. Ernst}.} \bibinfo{year}{2014}\natexlab{}.
\newblock \showarticletitle{{Defects4J}: A {Database} of existing faults to enable controlled testing studies for {Java} programs}. In \bibinfo{booktitle}{\emph{ISSTA 2014, Proceedings of the 2014 International Symposium on Software Testing and Analysis}}. \bibinfo{address}{San Jose, CA, USA}, \bibinfo{pages}{437--440}.
\newblock
\href{https://doi.org/10.1145/2610384.2628055}{doi:\nolinkurl{10.1145/2610384.2628055}}
\newblock
\shownote{Tool demo}.


\bibitem[Kästner(2022)]%
        {msr_keynote}
\bibfield{author}{\bibinfo{person}{Christian Kästner}.} \bibinfo{year}{2022}\natexlab{}.
\newblock \bibinfo{title}{From Models to Systems: Rethinking the Role of Software Engineering for ML}.
\newblock \bibinfo{howpublished}{\textsc{url:}~\url{https://www.youtube.com/watch?v=_m-m90S_4Gg}}.
\newblock


\bibitem[Le et~al\mbox{.}(2021)]%
        {le2021deepcva}
\bibfield{author}{\bibinfo{person}{Triet Huynh~Minh Le}, \bibinfo{person}{David Hin}, \bibinfo{person}{Roland Croft}, {and} \bibinfo{person}{M~Ali Babar}.} \bibinfo{year}{2021}\natexlab{}.
\newblock \showarticletitle{Deepcva: Automated commit-level vulnerability assessment with deep multi-task learning}. In \bibinfo{booktitle}{\emph{2021 36th IEEE/ACM International Conference on Automated Software Engineering (ASE)}}. IEEE, \bibinfo{pages}{717--729}.
\newblock
\href{https://doi.org/10.1109/ASE51524.2021.9678622}{doi:\nolinkurl{10.1109/ASE51524.2021.9678622}}


\bibitem[Levin and Yehudai(2017)]%
        {levin2017boosting}
\bibfield{author}{\bibinfo{person}{Stanislav Levin} {and} \bibinfo{person}{Amiram Yehudai}.} \bibinfo{year}{2017}\natexlab{}.
\newblock \showarticletitle{Boosting automatic commit classification into maintenance activities by utilizing source code changes}. In \bibinfo{booktitle}{\emph{Proceedings of the 13th International Conference on Predictive Models and Data Analytics in Software Engineering}}. \bibinfo{pages}{97--106}.
\newblock
\href{https://doi.org/10.1145/3127005.3127016}{doi:\nolinkurl{10.1145/3127005.3127016}}


\bibitem[Li et~al\mbox{.}(2024)]%
        {li2024effectiveness}
\bibfield{author}{\bibinfo{person}{Zhen Li}, \bibinfo{person}{Ning Wang}, \bibinfo{person}{Deqing Zou}, \bibinfo{person}{Yating Li}, \bibinfo{person}{Ruqian Zhang}, \bibinfo{person}{Shouhuai Xu}, \bibinfo{person}{Chao Zhang}, {and} \bibinfo{person}{Hai Jin}.} \bibinfo{year}{2024}\natexlab{}.
\newblock \showarticletitle{On the Effectiveness of Function-Level Vulnerability Detectors for Inter-Procedural Vulnerabilities}. In \bibinfo{booktitle}{\emph{Proceedings of the IEEE/ACM 46th International Conference on Software Engineering}} (Lisbon, Portugal) \emph{(\bibinfo{series}{ICSE '24})}. \bibinfo{publisher}{Association for Computing Machinery}, \bibinfo{address}{New York, NY, USA}, Article \bibinfo{articleno}{157}, \bibinfo{numpages}{12}~pages.
\newblock
\showISBNx{9798400702174}
\href{https://doi.org/10.1145/3597503.3639218}{doi:\nolinkurl{10.1145/3597503.3639218}}


\bibitem[Li et~al\mbox{.}(2019)]%
        {li2019comparative}
\bibfield{author}{\bibinfo{person}{Zhen Li}, \bibinfo{person}{Deqing Zou}, \bibinfo{person}{Jing Tang}, \bibinfo{person}{Zhihao Zhang}, \bibinfo{person}{Mingqian Sun}, {and} \bibinfo{person}{Hai Jin}.} \bibinfo{year}{2019}\natexlab{}.
\newblock \showarticletitle{A comparative study of deep learning-based vulnerability detection system}.
\newblock \bibinfo{journal}{\emph{IEEE Access}}  \bibinfo{volume}{7} (\bibinfo{year}{2019}), \bibinfo{pages}{103184--103197}.
\newblock
\href{https://doi.org/10.1109/ACCESS.2019.2930578}{doi:\nolinkurl{10.1109/ACCESS.2019.2930578}}


\bibitem[Li et~al\mbox{.}(2022a)]%
        {li_sysevr_2022}
\bibfield{author}{\bibinfo{person}{Zhen Li}, \bibinfo{person}{Deqing Zou}, \bibinfo{person}{Shouhuai Xu}, \bibinfo{person}{Hai Jin}, \bibinfo{person}{Yawei Zhu}, {and} \bibinfo{person}{Zhaoxuan Chen}.} \bibinfo{year}{2022}\natexlab{a}.
\newblock \showarticletitle{SySeVR: A Framework for Using Deep Learning to Detect Software Vulnerabilities}.
\newblock \bibinfo{journal}{\emph{IEEE Transactions on Dependable and Secure Computing}} \bibinfo{volume}{19}, \bibinfo{number}{4} (\bibinfo{date}{July} \bibinfo{year}{2022}), \bibinfo{pages}{2244–2258}.
\newblock
\showISSN{2160-9209}
\href{https://doi.org/10.1109/tdsc.2021.3051525}{doi:\nolinkurl{10.1109/tdsc.2021.3051525}}


\bibitem[Li et~al\mbox{.}(2018)]%
        {li_vuldeepecker_2018}
\bibfield{author}{\bibinfo{person}{Zhen Li}, \bibinfo{person}{Deqing Zou}, \bibinfo{person}{Shouhuai Xu}, \bibinfo{person}{Xinyu Ou}, \bibinfo{person}{Hai Jin}, \bibinfo{person}{Sujuan Wang}, \bibinfo{person}{Zhijun Deng}, {and} \bibinfo{person}{Yuyi Zhong}.} \bibinfo{year}{2018}\natexlab{}.
\newblock \showarticletitle{{VulDeePecker}: A Deep Learning-Based System for Vulnerability Detection}. In \bibinfo{booktitle}{\emph{Proceedings 2018 Network and Distributed System Security Symposium}}.
\newblock
\href{https://doi.org/10.14722/ndss.2018.23158}{doi:\nolinkurl{10.14722/ndss.2018.23158}}
\showeprint[arxiv]{1801.01681 [cs]}


\bibitem[Li et~al\mbox{.}(2022b)]%
        {vdp-data}
\bibfield{author}{\bibinfo{person}{Zhen Li}, \bibinfo{person}{Deqing Zou}, \bibinfo{person}{Shouhuai Xu}, \bibinfo{person}{Xinyu Ou}, \bibinfo{person}{Hai Jin}, \bibinfo{person}{Sujuan Wang}, \bibinfo{person}{Zhijun Deng}, {and} \bibinfo{person}{Yuyi Zhong}.} \bibinfo{year}{2022}\natexlab{b}.
\newblock \bibinfo{title}{VulDeePecker}.
\newblock \bibinfo{howpublished}{\textsc{url:}~\url{https://github.com/CGCL-codes/VulDeePecker}}.
\newblock


\bibitem[Liu et~al\mbox{.}(2019)]%
        {liu2019roberta}
\bibfield{author}{\bibinfo{person}{Yinhan Liu}, \bibinfo{person}{Myle Ott}, \bibinfo{person}{Naman Goyal}, \bibinfo{person}{Jingfei Du}, \bibinfo{person}{Mandar Joshi}, \bibinfo{person}{Danqi Chen}, \bibinfo{person}{Omer Levy}, \bibinfo{person}{Mike Lewis}, \bibinfo{person}{Luke Zettlemoyer}, {and} \bibinfo{person}{Veselin Stoyanov}.} \bibinfo{year}{2019}\natexlab{}.
\newblock \bibinfo{title}{RoBERTa: A Robustly Optimized BERT Pretraining Approach}.
\newblock
\href{https://doi.org/10.48550/arXiv.1907.11692}{doi:\nolinkurl{10.48550/arXiv.1907.11692}}
\showeprint[arxiv]{1907.11692}~[cs.CL]


\bibitem[Lwakatare et~al\mbox{.}(2020)]%
        {lwakatare2020large}
\bibfield{author}{\bibinfo{person}{Lucy~Ellen Lwakatare}, \bibinfo{person}{Aiswarya Raj}, \bibinfo{person}{Ivica Crnkovic}, \bibinfo{person}{Jan Bosch}, {and} \bibinfo{person}{Helena~Holmstr{\"o}m Olsson}.} \bibinfo{year}{2020}\natexlab{}.
\newblock \showarticletitle{Large-scale machine learning systems in real-world industrial settings: A review of challenges and solutions}.
\newblock \bibinfo{journal}{\emph{Information and software technology}}  \bibinfo{volume}{127} (\bibinfo{year}{2020}), \bibinfo{pages}{106368}.
\newblock
\href{https://doi.org/10.1016/j.infsof.2020.106368}{doi:\nolinkurl{10.1016/j.infsof.2020.106368}}


\bibitem[Mamede et~al\mbox{.}(2023)]%
        {vdet}
\bibfield{author}{\bibinfo{person}{Cl\'{a}udia Mamede}, \bibinfo{person}{Eduard Pinconschi}, {and} \bibinfo{person}{Rui Abreu}.} \bibinfo{year}{2023}\natexlab{}.
\newblock \showarticletitle{A transformer-based IDE plugin for vulnerability detection}. In \bibinfo{booktitle}{\emph{Proceedings of the 37th IEEE/ACM International Conference on Automated Software Engineering}} (Rochester, MI, USA) \emph{(\bibinfo{series}{ASE '22})}. \bibinfo{publisher}{Association for Computing Machinery}, \bibinfo{address}{New York, NY, USA}, Article \bibinfo{articleno}{149}, \bibinfo{numpages}{4}~pages.
\newblock
\showISBNx{9781450394758}
\href{https://doi.org/10.1145/3551349.3559534}{doi:\nolinkurl{10.1145/3551349.3559534}}


\bibitem[Mao et~al\mbox{.}(2024)]%
        {mao2024towards}
\bibfield{author}{\bibinfo{person}{Qiheng Mao}, \bibinfo{person}{Zhenhao Li}, \bibinfo{person}{Xing Hu}, \bibinfo{person}{Kui Liu}, \bibinfo{person}{Xin Xia}, {and} \bibinfo{person}{Jianling Sun}.} \bibinfo{year}{2024}\natexlab{}.
\newblock \showarticletitle{Towards Effectively Detecting and Explaining Vulnerabilities Using Large Language Models}.
\newblock \bibinfo{journal}{\emph{arXiv preprint arXiv:2406.09701}} (\bibinfo{year}{2024}).
\newblock
\href{https://doi.org/10.48550/arXiv.2406.09701}{doi:\nolinkurl{10.48550/arXiv.2406.09701}}


\bibitem[Marjanov et~al\mbox{.}(2022)]%
        {marjanov2022machine}
\bibfield{author}{\bibinfo{person}{Tina Marjanov}, \bibinfo{person}{Ivan Pashchenko}, {and} \bibinfo{person}{Fabio Massacci}.} \bibinfo{year}{2022}\natexlab{}.
\newblock \showarticletitle{Machine Learning for Source Code Vulnerability Detection: What Works and What Isn’t There Yet}.
\newblock \bibinfo{journal}{\emph{IEEE Security \& Privacy}} \bibinfo{volume}{20}, \bibinfo{number}{5} (\bibinfo{year}{2022}), \bibinfo{pages}{60--76}.
\newblock
\href{https://doi.org/10.1109/MSEC.2022.3176058}{doi:\nolinkurl{10.1109/MSEC.2022.3176058}}


\bibitem[Massacci et~al\mbox{.}(2011)]%
        {massacci2011after}
\bibfield{author}{\bibinfo{person}{Fabio Massacci}, \bibinfo{person}{Stephan Neuhaus}, {and} \bibinfo{person}{Viet~Hung Nguyen}.} \bibinfo{year}{2011}\natexlab{}.
\newblock \showarticletitle{After-life vulnerabilities: a study on firefox evolution, its vulnerabilities, and fixes}. In \bibinfo{booktitle}{\emph{International Symposium on Engineering Secure Software and Systems}}. Springer, \bibinfo{pages}{195--208}.
\newblock
\href{https://doi.org/10.1007/978-3-642-19125-1_15}{doi:\nolinkurl{10.1007/978-3-642-19125-1_15}}


\bibitem[Mauczka et~al\mbox{.}(2015)]%
        {mauczka2015dataset}
\bibfield{author}{\bibinfo{person}{Andreas Mauczka}, \bibinfo{person}{Florian Brosch}, \bibinfo{person}{Christian Schanes}, {and} \bibinfo{person}{Thomas Grechenig}.} \bibinfo{year}{2015}\natexlab{}.
\newblock \showarticletitle{Dataset of developer-labeled commit messages}. In \bibinfo{booktitle}{\emph{2015 IEEE/ACM 12th Working Conference on Mining Software Repositories}}. IEEE, \bibinfo{pages}{490--493}.
\newblock
\href{https://doi.org/10.1109/MSR.2015.71}{doi:\nolinkurl{10.1109/MSR.2015.71}}


\bibitem[Mirsky et~al\mbox{.}(2023)]%
        {mirsky2023vulchecker}
\bibfield{author}{\bibinfo{person}{Yisroel Mirsky}, \bibinfo{person}{George Macon}, \bibinfo{person}{Michael Brown}, \bibinfo{person}{Carter Yagemann}, \bibinfo{person}{Matthew Pruett}, \bibinfo{person}{Evan Downing}, \bibinfo{person}{Sukarno Mertoguno}, {and} \bibinfo{person}{Wenke Lee}.} \bibinfo{year}{2023}\natexlab{}.
\newblock \showarticletitle{VulChecker: Graph-based Vulnerability Localization in Source Code}. In \bibinfo{booktitle}{\emph{31st USENIX Security Symposium, Security 2022}}.
\newblock
\urldef\tempurl%
\url{https://dl.acm.org/doi/abs/10.5555/3620237.3620604}
\showURL{%
\tempurl}


\bibitem[Nguyen et~al\mbox{.}(2022)]%
        {nguyen2022regvd}
\bibfield{author}{\bibinfo{person}{Van-Anh Nguyen}, \bibinfo{person}{Dai~Quoc Nguyen}, \bibinfo{person}{Van Nguyen}, \bibinfo{person}{Trung Le}, \bibinfo{person}{Quan~Hung Tran}, {and} \bibinfo{person}{Dinh Phung}.} \bibinfo{year}{2022}\natexlab{}.
\newblock \showarticletitle{ReGVD: Revisiting graph neural networks for vulnerability detection}. In \bibinfo{booktitle}{\emph{Proceedings of the ACM/IEEE 44th International Conference on Software Engineering: Companion Proceedings}}. \bibinfo{pages}{178--182}.
\newblock
\href{https://doi.org/10.1145/3510454.3516865}{doi:\nolinkurl{10.1145/3510454.3516865}}


\bibitem[Nguyen et~al\mbox{.}(2016)]%
        {nguyen2016automatic}
\bibfield{author}{\bibinfo{person}{Viet~Hung Nguyen}, \bibinfo{person}{Stanislav Dashevskyi}, {and} \bibinfo{person}{Fabio Massacci}.} \bibinfo{year}{2016}\natexlab{}.
\newblock \showarticletitle{An automatic method for assessing the versions affected by a vulnerability}.
\newblock \bibinfo{journal}{\emph{Empirical Software Engineering}}  \bibinfo{volume}{21} (\bibinfo{year}{2016}), \bibinfo{pages}{2268--2297}.
\newblock
\href{https://doi.org/10.1007/s10664-015-9408-2}{doi:\nolinkurl{10.1007/s10664-015-9408-2}}


\bibitem[Ni et~al\mbox{.}(2023)]%
        {svuld}
\bibfield{author}{\bibinfo{person}{Chao Ni}, \bibinfo{person}{Xin Yin}, \bibinfo{person}{Kaiwen Yang}, \bibinfo{person}{Dehai Zhao}, \bibinfo{person}{Zhenchang Xing}, {and} \bibinfo{person}{Xin Xia}.} \bibinfo{year}{2023}\natexlab{}.
\newblock \showarticletitle{Distinguishing Look-Alike Innocent and Vulnerable Code by Subtle Semantic Representation Learning and Explanation}. In \bibinfo{booktitle}{\emph{Proceedings of the 31st ACM Joint European Software Engineering Conference and Symposium on the Foundations of Software Engineering}} (San Francisco, CA, USA) \emph{(\bibinfo{series}{ESEC/FSE 2023})}. \bibinfo{publisher}{Association for Computing Machinery}, \bibinfo{address}{New York, NY, USA}, \bibinfo{pages}{1611–1622}.
\newblock
\showISBNx{9798400703270}
\href{https://doi.org/10.1145/3611643.3616358}{doi:\nolinkurl{10.1145/3611643.3616358}}


\bibitem[Nikitopoulos et~al\mbox{.}(2021)]%
        {nikitopoulos2021crossvul}
\bibfield{author}{\bibinfo{person}{Georgios Nikitopoulos}, \bibinfo{person}{Konstantina Dritsa}, \bibinfo{person}{Panos Louridas}, {and} \bibinfo{person}{Dimitris Mitropoulos}.} \bibinfo{year}{2021}\natexlab{}.
\newblock \showarticletitle{CrossVul: a cross-language vulnerability dataset with commit data}. In \bibinfo{booktitle}{\emph{Proceedings of the 29th ACM Joint Meeting on European Software Engineering Conference and Symposium on the Foundations of Software Engineering}}. \bibinfo{pages}{1565--1569}.
\newblock
\href{https://doi.org/10.1145/3468264.3473122}{doi:\nolinkurl{10.1145/3468264.3473122}}


\bibitem[NIST(2024a)]%
        {nvd}
\bibfield{author}{\bibinfo{person}{NIST}.} \bibinfo{year}{2024}\natexlab{a}.
\newblock \bibinfo{title}{National Vulnerability Database}.
\newblock
\urldef\tempurl%
\url{https://nvd.nist.gov/}
\showURL{%
\tempurl}


\bibitem[NIST(2024b)]%
        {nvd-api}
\bibfield{author}{\bibinfo{person}{NIST}.} \bibinfo{year}{2024}\natexlab{b}.
\newblock \bibinfo{title}{National Vulnerability Database: Vulnerability API}.
\newblock
\urldef\tempurl%
\url{https://nvd.nist.gov/developers/vulnerabilities}
\showURL{%
\tempurl}


\bibitem[NIST(2024c)]%
        {sard}
\bibfield{author}{\bibinfo{person}{NIST}.} \bibinfo{year}{2024}\natexlab{c}.
\newblock \bibinfo{title}{Software Assurance Reference Dataset}.
\newblock
\urldef\tempurl%
\url{https://samate.nist.gov/SRD/index.php}
\showURL{%
\tempurl}


\bibitem[Okun et~al\mbox{.}(2013)]%
        {okun2013report}
\bibfield{author}{\bibinfo{person}{Vadim Okun}, \bibinfo{person}{Aurelien Delaitre}, \bibinfo{person}{Paul~E Black}, {et~al\mbox{.}}} \bibinfo{year}{2013}\natexlab{}.
\newblock \showarticletitle{Report on the static analysis tool exposition (sate) iv}.
\newblock \bibinfo{journal}{\emph{NIST Special Publication}}  \bibinfo{volume}{500} (\bibinfo{year}{2013}), \bibinfo{pages}{297}.
\newblock
\href{https://doi.org/10.6028/NIST.SP.500-297}{doi:\nolinkurl{10.6028/NIST.SP.500-297}}


\bibitem[Ozment and Schechter(2006)]%
        {ozment2006milk}
\bibfield{author}{\bibinfo{person}{Andy Ozment} {and} \bibinfo{person}{Stuart~E Schechter}.} \bibinfo{year}{2006}\natexlab{}.
\newblock \showarticletitle{Milk or wine: does software security improve with age?}. In \bibinfo{booktitle}{\emph{USENIX Security Symposium}}, Vol.~\bibinfo{volume}{6}. \bibinfo{pages}{10--5555}.
\newblock
\href{https://doi.org/10.5555/1267336.1267343}{doi:\nolinkurl{10.5555/1267336.1267343}}


\bibitem[Pachouly et~al\mbox{.}(2022)]%
        {pachouly2022systematic}
\bibfield{author}{\bibinfo{person}{Jalaj Pachouly}, \bibinfo{person}{Swati Ahirrao}, \bibinfo{person}{Ketan Kotecha}, \bibinfo{person}{Ganeshsree Selvachandran}, {and} \bibinfo{person}{Ajith Abraham}.} \bibinfo{year}{2022}\natexlab{}.
\newblock \showarticletitle{A systematic literature review on software defect prediction using artificial intelligence: Datasets, Data Validation Methods, Approaches, and Tools}.
\newblock \bibinfo{journal}{\emph{Engineering Applications of Artificial Intelligence}}  \bibinfo{volume}{111} (\bibinfo{year}{2022}), \bibinfo{pages}{104773}.
\newblock
\href{https://doi.org/10.1016/j.engappai.2022.104773}{doi:\nolinkurl{10.1016/j.engappai.2022.104773}}


\bibitem[Pan et~al\mbox{.}(2023)]%
        {TreeVUL}
\bibfield{author}{\bibinfo{person}{Shengyi Pan}, \bibinfo{person}{Lingfeng Bao}, \bibinfo{person}{Xin Xia}, \bibinfo{person}{David Lo}, {and} \bibinfo{person}{Shanping Li}.} \bibinfo{year}{2023}\natexlab{}.
\newblock \showarticletitle{Fine-grained Commit-level Vulnerability Type Prediction by CWE Tree Structure}. In \bibinfo{booktitle}{\emph{2023 IEEE/ACM 45th International Conference on Software Engineering (ICSE)}}. \bibinfo{pages}{957--969}.
\newblock
\href{https://doi.org/10.1109/ICSE48619.2023.00088}{doi:\nolinkurl{10.1109/ICSE48619.2023.00088}}


\bibitem[Paramitha et~al\mbox{.}(2025a)]%
        {zenodo-us-1}
\bibfield{author}{\bibinfo{person}{Ranindya Paramitha}, \bibinfo{person}{Yuan Feng}, {and} \bibinfo{person}{Fabio Massacci}.} \bibinfo{year}{2025}\natexlab{a}.
\newblock \bibinfo{title}{Replication package on Zenodo Part 1 (NVD Vuldeepecker Dataset)}.
\newblock \bibinfo{howpublished}{\href{https://doi.org/10.5281/zenodo.8207883}{Zenodo link}. If the link does not work, copy and paste the following link \url{https://doi.org/10.5281/zenodo.8207883}}.
\newblock


\bibitem[Paramitha et~al\mbox{.}(2025b)]%
        {zenodo-us-2}
\bibfield{author}{\bibinfo{person}{Ranindya Paramitha}, \bibinfo{person}{Yuan Feng}, {and} \bibinfo{person}{Fabio Massacci}.} \bibinfo{year}{2025}\natexlab{b}.
\newblock \bibinfo{title}{Replication package on Zenodo Part 2 (LINUX Dataset)}.
\newblock \bibinfo{howpublished}{\href{https://doi.org/10.5281/zenodo.10960662}{Zenodo link}. If the link does not work, copy and paste the link in the following link \url{https://doi.org/10.5281/zenodo.10960662}}.
\newblock


\bibitem[Paramitha et~al\mbox{.}(2025c)]%
        {zenodo-us-3}
\bibfield{author}{\bibinfo{person}{Ranindya Paramitha}, \bibinfo{person}{Yuan Feng}, {and} \bibinfo{person}{Fabio Massacci}.} \bibinfo{year}{2025}\natexlab{c}.
\newblock \bibinfo{title}{Replication package on Zenodo Part 3 (OPENSSL Dataset)}.
\newblock \bibinfo{howpublished}{\href{https://doi.org/10.5281/zenodo.10966117}{Zenodo link}. If the link does not work, copy and paste the link in the following link \url{https://doi.org/10.5281/zenodo.10966117}}.
\newblock


\bibitem[Paramitha et~al\mbox{.}(2025d)]%
        {zenodo-us-4}
\bibfield{author}{\bibinfo{person}{Ranindya Paramitha}, \bibinfo{person}{Yuan Feng}, {and} \bibinfo{person}{Fabio Massacci}.} \bibinfo{year}{2025}\natexlab{d}.
\newblock \bibinfo{title}{Replication package on Zenodo Part 4 (POPPLER Dataset)}.
\newblock \bibinfo{howpublished}{\href{https://doi.org/10.5281/zenodo.14713143}{Zenodo link}. If the link does not work, copy and paste the following link \url{https://doi.org/10.5281/zenodo.14713143}}.
\newblock


\bibitem[Perl et~al\mbox{.}(2015)]%
        {perl2015vccfinder}
\bibfield{author}{\bibinfo{person}{Henning Perl}, \bibinfo{person}{Sergej Dechand}, \bibinfo{person}{Matthew Smith}, \bibinfo{person}{Daniel Arp}, \bibinfo{person}{Fabian Yamaguchi}, \bibinfo{person}{Konrad Rieck}, \bibinfo{person}{Sascha Fahl}, {and} \bibinfo{person}{Yasemin Acar}.} \bibinfo{year}{2015}\natexlab{}.
\newblock \showarticletitle{Vccfinder: Finding potential vulnerabilities in open-source projects to assist code audits}. In \bibinfo{booktitle}{\emph{Proceedings of the 22nd ACM SIGSAC Conference on Computer and Communications Security}}. \bibinfo{pages}{426--437}.
\newblock
\href{https://doi.org/10.1145/2810103.2813604}{doi:\nolinkurl{10.1145/2810103.2813604}}


\bibitem[Ponta et~al\mbox{.}(2019)]%
        {ponta2019msr}
\bibfield{author}{\bibinfo{person}{Serena~E. Ponta}, \bibinfo{person}{Henrik Plate}, \bibinfo{person}{Antonino Sabetta}, \bibinfo{person}{Michele Bezzi}, {and} \bibinfo{person}{C´edric Dangremont}.} \bibinfo{year}{2019}\natexlab{}.
\newblock \showarticletitle{A Manually-Curated Dataset of Fixes to Vulnerabilities of Open-Source Software}. In \bibinfo{booktitle}{\emph{Proceedings of the 16th International Conference on Mining Software Repositories}}.
\newblock
\href{https://doi.org/10.1109/MSR.2019.00064}{doi:\nolinkurl{10.1109/MSR.2019.00064}}


\bibitem[Risse and B{\"o}hme(2024)]%
        {risse2024uncovering}
\bibfield{author}{\bibinfo{person}{Niklas Risse} {and} \bibinfo{person}{Marcel B{\"o}hme}.} \bibinfo{year}{2024}\natexlab{}.
\newblock \showarticletitle{Uncovering the Limits of Machine Learning for Automatic Vulnerability Detection}. In \bibinfo{booktitle}{\emph{USENIX Security Symposium 2024}}. \bibinfo{pages}{19}.
\newblock
\href{https://doi.org/10.5555/3698900.3699138}{doi:\nolinkurl{10.5555/3698900.3699138}}


\bibitem[Ruscio and Gera(2013)]%
        {ruscio2013generalizations}
\bibfield{author}{\bibinfo{person}{John Ruscio} {and} \bibinfo{person}{Benjamin~Lee Gera}.} \bibinfo{year}{2013}\natexlab{}.
\newblock \showarticletitle{Generalizations and extensions of the probability of superiority effect size estimator}.
\newblock \bibinfo{journal}{\emph{Multivariate behavioral research}} \bibinfo{volume}{48}, \bibinfo{number}{2} (\bibinfo{year}{2013}), \bibinfo{pages}{208--219}.
\newblock
\href{https://doi.org/10.1080/00273171.2012.738184}{doi:\nolinkurl{10.1080/00273171.2012.738184}}


\bibitem[Russell et~al\mbox{.}(2018)]%
        {russell2018automated}
\bibfield{author}{\bibinfo{person}{Rebecca Russell}, \bibinfo{person}{Louis Kim}, \bibinfo{person}{Lei Hamilton}, \bibinfo{person}{Tomo Lazovich}, \bibinfo{person}{Jacob Harer}, \bibinfo{person}{Onur Ozdemir}, \bibinfo{person}{Paul Ellingwood}, {and} \bibinfo{person}{Marc McConley}.} \bibinfo{year}{2018}\natexlab{}.
\newblock \showarticletitle{Automated vulnerability detection in source code using deep representation learning}. In \bibinfo{booktitle}{\emph{2018 17th IEEE international conference on machine learning and applications (ICMLA)}}. IEEE, \bibinfo{pages}{757--762}.
\newblock
\href{https://doi.org/10.1109/CCWC60891.2024.10427574}{doi:\nolinkurl{10.1109/CCWC60891.2024.10427574}}


\bibitem[Safdari(2018)]%
        {safdari2018multiclass}
\bibfield{author}{\bibinfo{person}{Nasir Safdari}.} \bibinfo{year}{2018}\natexlab{}.
\newblock \bibinfo{title}{{Multi-Class-Text-Classification----Random-Forest}}.
\newblock \bibinfo{howpublished}{Github}.
\newblock
\urldef\tempurl%
\url{https://github.com/nxs5899/Multi-Class-Text-Classification----Random-Forest}
\showURL{%
\tempurl}


\bibitem[Sayyad\-Shirabad and Menzies(2005)]%
        {promise}
\bibfield{author}{\bibinfo{person}{J. Sayyad\-Shirabad} {and} \bibinfo{person}{T.J. Menzies}.} \bibinfo{year}{2005}\natexlab{}.
\newblock \bibinfo{title}{{The {PROMISE} Repository of Software Engineering Databases.}}
\newblock \bibinfo{howpublished}{School of Information Technology and Engineering, University of Ottawa, Canada}.
\newblock
\urldef\tempurl%
\url{http://promise.site.uottawa.ca/SERepository}
\showURL{%
\tempurl}


\bibitem[Scandariato et~al\mbox{.}(2014)]%
        {Scandariato}
\bibfield{author}{\bibinfo{person}{Riccardo Scandariato}, \bibinfo{person}{James Walden}, \bibinfo{person}{Aram Hovsepyan}, {and} \bibinfo{person}{Wouter Joosen}.} \bibinfo{year}{2014}\natexlab{}.
\newblock \showarticletitle{Predicting Vulnerable Software Components via Text Mining}.
\newblock \bibinfo{journal}{\emph{IEEE Transactions on Software Engineering}} \bibinfo{volume}{40}, \bibinfo{number}{10} (\bibinfo{year}{2014}), \bibinfo{pages}{993--1006}.
\newblock
\href{https://doi.org/10.1109/TSE.2014.2340398}{doi:\nolinkurl{10.1109/TSE.2014.2340398}}


\bibitem[Shepperd et~al\mbox{.}(2013)]%
        {shepperd2013data}
\bibfield{author}{\bibinfo{person}{Martin Shepperd}, \bibinfo{person}{Qinbao Song}, \bibinfo{person}{Zhongbin Sun}, {and} \bibinfo{person}{Carolyn Mair}.} \bibinfo{year}{2013}\natexlab{}.
\newblock \showarticletitle{Data quality: Some comments on the nasa software defect datasets}.
\newblock \bibinfo{journal}{\emph{IEEE Transactions on software engineering}} \bibinfo{volume}{39}, \bibinfo{number}{9} (\bibinfo{year}{2013}), \bibinfo{pages}{1208--1215}.
\newblock
\href{https://doi.org/10.1109/TSE.2013.11}{doi:\nolinkurl{10.1109/TSE.2013.11}}


\bibitem[Shin et~al\mbox{.}(2011)]%
        {Shin_eval}
\bibfield{author}{\bibinfo{person}{Yonghee Shin}, \bibinfo{person}{Andrew Meneely}, \bibinfo{person}{Laurie Williams}, {and} \bibinfo{person}{Jason Osborne}.} \bibinfo{year}{2011}\natexlab{}.
\newblock \showarticletitle{Evaluating Complexity, Code Churn, and Developer Activity Metrics as Indicators of Software Vulnerabilities}.
\newblock \bibinfo{journal}{\emph{IEEE Trans. Software Eng.}}  \bibinfo{volume}{37} (\bibinfo{date}{11} \bibinfo{year}{2011}), \bibinfo{pages}{772--787}.
\newblock
\href{https://doi.org/10.1109/TSE.2010.81}{doi:\nolinkurl{10.1109/TSE.2010.81}}


\bibitem[Steenhoek et~al\mbox{.}(2023)]%
        {steenhoek2023empirical}
\bibfield{author}{\bibinfo{person}{Benjamin Steenhoek}, \bibinfo{person}{Md~Mahbubur Rahman}, \bibinfo{person}{Richard Jiles}, {and} \bibinfo{person}{Wei Le}.} \bibinfo{year}{2023}\natexlab{}.
\newblock \showarticletitle{An empirical study of deep learning models for vulnerability detection}. In \bibinfo{booktitle}{\emph{2023 IEEE/ACM 45th International Conference on Software Engineering (ICSE)}}. IEEE, \bibinfo{pages}{2237--2248}.
\newblock
\href{https://doi.org/10.1109/ICSE48619.2023.00188}{doi:\nolinkurl{10.1109/ICSE48619.2023.00188}}


\bibitem[Yatish et~al\mbox{.}(2019)]%
        {yatish2019mining}
\bibfield{author}{\bibinfo{person}{Suraj Yatish}, \bibinfo{person}{Jirayus Jiarpakdee}, \bibinfo{person}{Patanamon Thongtanunam}, {and} \bibinfo{person}{Chakkrit Tantithamthavorn}.} \bibinfo{year}{2019}\natexlab{}.
\newblock \showarticletitle{Mining software defects: Should we consider affected releases?}. In \bibinfo{booktitle}{\emph{2019 IEEE/ACM 41st International Conference on Software Engineering (ICSE)}}. IEEE, \bibinfo{pages}{654--665}.
\newblock
\href{https://doi.org/10.1109/ICSE.2019.00075}{doi:\nolinkurl{10.1109/ICSE.2019.00075}}


\bibitem[Yuan et~al\mbox{.}(2023)]%
        {vulbg}
\bibfield{author}{\bibinfo{person}{Bin Yuan}, \bibinfo{person}{Yifan Lu}, \bibinfo{person}{Yilin Fang}, \bibinfo{person}{Yueming Wu}, \bibinfo{person}{Deqing Zou}, \bibinfo{person}{Zhen Li}, \bibinfo{person}{Zhi Li}, {and} \bibinfo{person}{Hai Jin}.} \bibinfo{year}{2023}\natexlab{}.
\newblock \showarticletitle{Enhancing Deep Learning-based Vulnerability Detection by Building Behavior Graph Model}. In \bibinfo{booktitle}{\emph{2023 IEEE/ACM 45th International Conference on Software Engineering (ICSE)}}. \bibinfo{pages}{2262--2274}.
\newblock
\href{https://doi.org/10.1109/ICSE48619.2023.00190}{doi:\nolinkurl{10.1109/ICSE48619.2023.00190}}


\bibitem[Zheng et~al\mbox{.}(2021)]%
        {zheng2021d2a}
\bibfield{author}{\bibinfo{person}{Yunhui Zheng}, \bibinfo{person}{Saurabh Pujar}, \bibinfo{person}{Burn Lewis}, \bibinfo{person}{Luca Buratti}, \bibinfo{person}{Edward Epstein}, \bibinfo{person}{Bo Yang}, \bibinfo{person}{Jim Laredo}, \bibinfo{person}{Alessandro Morari}, {and} \bibinfo{person}{Zhong Su}.} \bibinfo{year}{2021}\natexlab{}.
\newblock \showarticletitle{D2a: A dataset built for ai-based vulnerability detection methods using differential analysis}. In \bibinfo{booktitle}{\emph{2021 IEEE/ACM 43rd International Conference on Software Engineering: Software Engineering in Practice (ICSE-SEIP)}}. IEEE, \bibinfo{pages}{111--120}.
\newblock
\href{https://doi.org/10.1109/ICSE-SEIP52600.2021.00020}{doi:\nolinkurl{10.1109/ICSE-SEIP52600.2021.00020}}


\bibitem[Zhou et~al\mbox{.}(2019)]%
        {zhou2019devign}
\bibfield{author}{\bibinfo{person}{Yaqin Zhou}, \bibinfo{person}{Shangqing Liu}, \bibinfo{person}{Jingkai Siow}, \bibinfo{person}{Xiaoning Du}, {and} \bibinfo{person}{Yang Liu}.} \bibinfo{year}{2019}\natexlab{}.
\newblock \showarticletitle{Devign: Effective vulnerability identification by learning comprehensive program semantics via graph neural networks}. In \bibinfo{booktitle}{\emph{Advances in Neural Information Processing Systems}}. \bibinfo{pages}{10197--10207}.
\newblock
\href{https://doi.org/10.5555/3454287.3455202}{doi:\nolinkurl{10.5555/3454287.3455202}}


\end{thebibliography}

\end{document}